# PMB-NN: Physiology-Centred Hybrid AI for Personalized Hemodynamic Monitoring from Photoplethysmography

Yaowen Zhang[a], Libera Fresiello[b], Peter H. Veltink[a], Dirk W. Donker[b,c] and Ying Wang[a,*]

[a]*Department of Biomedical Signals and Systems, University of Twente, Enschede, 7522NB, The Netherlands*
[b]*Department of Cardiovascular and Respiratory Physiology, University of Twente, Enschede, 7522NB, The Netherlands*
[c]*Department of Intensive Care, University Medical Center Utrecht, Utrecht, 3584CX, The Netherlands*

ABSTRACT

Background and Objective: Continuous monitoring of blood pressure (BP) and hemodynamic parameters such as peripheral resistance ($R$) and arterial compliance ($C$) are critical for early vascular dysfunction detection and therapy optimization. While photoplethysmography (PPG) wearables has gained widespread popularity, existing data-driven methods for BP estimation lack physiological interpretability.

*Methods:* We advanced our previously proposed physiology-centered hybrid AI method—the Physiological Model-Based Neural Network (PMB-NN)—in arterial hemodynamic monitoring, that unifies deep learning with a 2-element Windkessel based model parameterized by $R$ and $C$ acting as physics constraints. The PMB-NN model was trained in a subject-specific manner using only PPG-derived timing features, while demographic information was used to infer an intermediate physiological variable: cardiac output. The model outputs personalized systolic and diastolic BP. We validated the model on 10 healthy adults performing static and cycling activities across two days for model's day-to-day robustness, benchmarked against state-of-the-art deep learning (DL) models (FCNN, CNN-LSTM, Transformer) and standalone Windkessel based physiological model (PM). Validation was conducted on three perspectives: accuracy, physiological interpretability and plausibility.

*Results:* PMB-NN achieved systolic BP accuracy (median MAE: 7.2 mmHg) comparable to deep learning benchmarks, despite yielding diastolic performance (median MAE: 3.9 mmHg) lower than DL models. Crucially, however, PMB-NN exhibited substantially higher physiological plausibility than both DL baselines and PM, suggesting that the hybrid architecture unifies and enhances the respective merits of physiological principles and data-driven techniques. Beyond BP, PMB-NN also identified $R$ (mean error: 0.15 mmHg·s/ml) and $C$ (mean error: -0.35 ml/mmHg) during training with accuracy similar to the standalone physiological model, demonstrating that the embedded physiological constraints confer interpretability to the hybrid AI framework.

*Conclusions:* These results position PMB-NN as a balanced, physiologically grounded alternative to purely data-driven approaches for daily hemodynamic monitoring.

## 1. Introduction

Hypertension remains one of the most pervasive global health challenges, affecting more than 1.28 billion adults worldwide, yet fewer than one in five patients achieve adequate blood pressure (BP) control [1]. Poor BP control accelerates vascular remodeling and arterial stiffening, further creates a vicious cycle that further exacerbates hypertension. A growing body of evidence indicates that arterial stiffening, which is not only a consequence but also a driver of elevated blood pressure, contributes to hypertension through mechanisms such as elastin degradation, collagen deposition, and smooth muscle cell dysfunction, ultimately increasing total peripheral resistance ($R$) and reducing arterial compliance ($C$) [2, 3]. These vascular changes promote left ventricular hypertrophy [4], impair baroreflex sensitivity [5], and exacerbate systolic hypertension, particularly in older adults [6]. Interplay between BP, $R$ and $C$ during physical activities is a critical oversight given that exercise-induced hemodynamic shifts exacerbate cardiovascular risk in hypertensive patients [7]. Consequently, daily continuous monitoring of BP alongside $R$ and $C$ is essential to capture the critical window for early detection of vascular dysfunction [8],[9] and timely optimization of antihypertensive therapy [10], and effective blood pressure control [11].

Recent advances in photoplethysmography (PPG)-based BP sensing have emerged as a promising alternative, leveraging non-invasive optical measurements of microvascular blood volume changes [12],[13]. Unlike conventional cuff-based measurements, which are intermittent and can disrupt activities or cause discomfort, PPG leverages non-invasive optical detection of microvascular blood volume changes to estimate BP unobtrusively and continuously [14],[15]. Single-PPG-sensor methodologies for BP estimation have been developed, bifurcating into parametric and non-parametric paradigms [16], offering simpler and more practical solutions for daily life monitoring than conventional pulse wave velocity (PWV) methods [17],[18],[19],[20]. Parametric models, rooted in physiological principles such as the Windkessel-based model [21],[22], utilize predefined hemodynamic equations to derive BP from features like systolic upstroke time and diastolic time [22]. While these models offer interpretability by linking PPG features to underlying cardiovascular mechanics, their reliance on high-quality signals and individualized calibration restricts utility in real-world settings where motion artifacts and ambient noise degrade signal fidelity [23],[24]. Conversely, non-parametric approaches including deep neural networks [25], convolutional architectures [26], and transformer-based models [27], leverage data-driven feature extraction to bypass explicit physiological assumptions. These methods demonstrate superior adaptability to noisy environments but suffer from 'black-box' limitations, obscuring clinical interpretability and hindering trust in critical healthcare applications [28]. The dichotomy between parametric and non-parametric models reveals a fundamental trade-off: physiological interpretability versus estimation robustness [29],[30]. Hybrid frameworks that synergize parametric and non-parametric approaches have the potential to theoretically balance physiological interpretability versus estimation robustness [31],[32]. For instance, one type of the hybrid AI frameworks: physics-informed neural networks (PINNs) that integrate mathematical models with neural networks, have been demonstrated to provide robust hemodynamic parameter estimation under conditions of limited data or incomplete system information, as reported in bioimpedance-based studies [33] and MRI-based studies [34],[25].

*Corresponding author
✉ y.zhang-12@utwente.nl (Y. Zhang); l.fresiello@utwente.nl (L. Fresiello); p.h.veltink@utwente.nl (P.H. Veltink); d.w.donker@utwente.nl (D.W. Donker); imwywk@gmail.com (Y. Wang)
ORCID(s):

This study builds upon our previously proposed Physiological Model–Based Neural Network (PMB-NN) framework [32], which was originally introduced using a physiology constrained neural network to embed a metabolic-cardiac regulation context for $\dot{V}O_2$ driven heart rate estimation. While the prior work embedded a simplified metabolic–cardiac model centered on heart rate dynamics, the present study represents a distinct different physiological subsystem—arterial hemodynamics—through embedding a Windkessel-based BP estimation model. By extending PMB-NN from metabolic–cardiac regulation to arterial hemodynamics, this work illustrates how the hybrid architecture can be instantiated across physiological subsystems governed by different mechanisms.

A preliminary version of this work was presented at 47th Annual International Conference of the IEEE Engineering in Medicine and Biology Society [35]. The present manuscript substantially extends that work through the incorporation of a cardiac output estimation framework, the establishment of a comprehensive evaluation dataset, and the deployment of additional assessment methodologies. This study aims to evaluate the proposed PMB-NN scheme through the following objectives:

1. Performance evaluation: To assess whether PMB-NN achieves higher blood pressure estimation accuracy than the standalone parametric model (PM) in BP estimation accuracy while maintaining competitive performance comparable to benchmark deep learning models: FCNN, CNN-LSTM, Transformer.
2. Physiological interpretability validation: To determine whether the model infers *R* and *C* parameters that remain physiologically interpretable and consistent across diverse activity conditions.
3. Physiological plausibility validation: To examine how the integration of physiological modeling with data-driven learning contributes to enhanced physiological plausibility of the hybrid AI scheme.

## 2. Method
### 2.1. Data acquisition

The study protocol was approved by the Ethics Committee for Computer and Information Science (EC-CIS) of the University of Twente (Approval No. 240831) and conducted at Roessingh Research and Development (RRD). 10 healthy subjects were enrolled after providing written informed consent. The study adhered to the principles of the Declaration of Helsinki.

Each subject completed two measurement sessions on separate days [median interval: 110 (20, 126) days] to provide repeated datasets required for machine learning training and testing per individuals, while also capturing day-to-day variability and reducing subject burden. The experimental protocol was identical on both days. The experimental protocol comprised three sequential physical activity-level phases conducted over a 35-minute period shown in Fig. 1: Static Posture (SP; 5 minutes seated followed by 5 minutes standing), Low-Intensity Cycling (LIC; 10 minutes at 45 rpm and 50 W) following a 1-minute preparation interval, and Moderate-Intensity Cycling (MIC; 10 minutes at 45 rpm and 100 W) following a 3-minute recovery interval. In total, there were 60 data segments excluding the preparation and recovery interval (2 days × 3 phrases × 10 subjects) collected from the 10 subjects on the two days. Exercise intensities were carefully selected to maintain subjects within their aerobic threshold. 10-minute duration for each cycling phase was selected to ensure sufficient time for hemodynamic stabilization of blood pressure, total peripheral resistance, and arterial compliance, thereby minimizing transient physiological fluctuations associated with exercise onset. The cycling protocol was performed on a standardized Bremshey® ergometer to ensure consistent mechanical loading across subjects.

Prior to the measurement, subjects were instructed to abstain from food for at least one hour to minimize gastrointestinal activity and ensure stable cardiovascular baseline readings. Subject characteristics including sex, age, height and weight were recorded. To minimize measurement artifacts, subjects acclimatized to the laboratory environment until hand temperature stabilized to normothermic levels, as cold-induced vasoconstriction could impair signal quality. The left index and middle fingers were cleansed with alcohol wipes to remove cutaneous oils and sweat, ensuring optimal sensor adhesion.

Physiological measurements were acquired using two systems: 64 Hz PPG signals were obtained from the left index finger using a Shimmer 3 GSR+ Unit (Shimmer Wearable Sensory Technology), while continuous blood pressure was monitored through volume-clamp finger cuff (Finometer® Model-2, Finapres Medical Systems) applied to the left middle finger. To ensure measurement accuracy, the left hand was maintained at heart level throughout all experimental phases using a specially designed sling, effectively eliminating potential hydrostatic pressure artifacts.

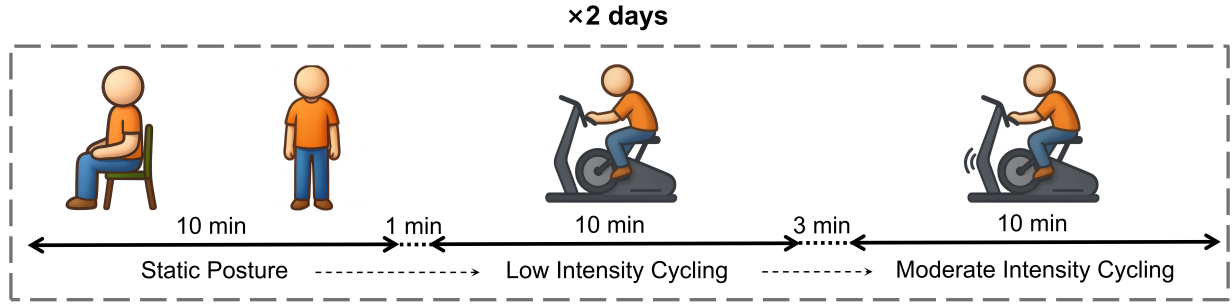

Figure 1: Experiment Protocol for data collection.

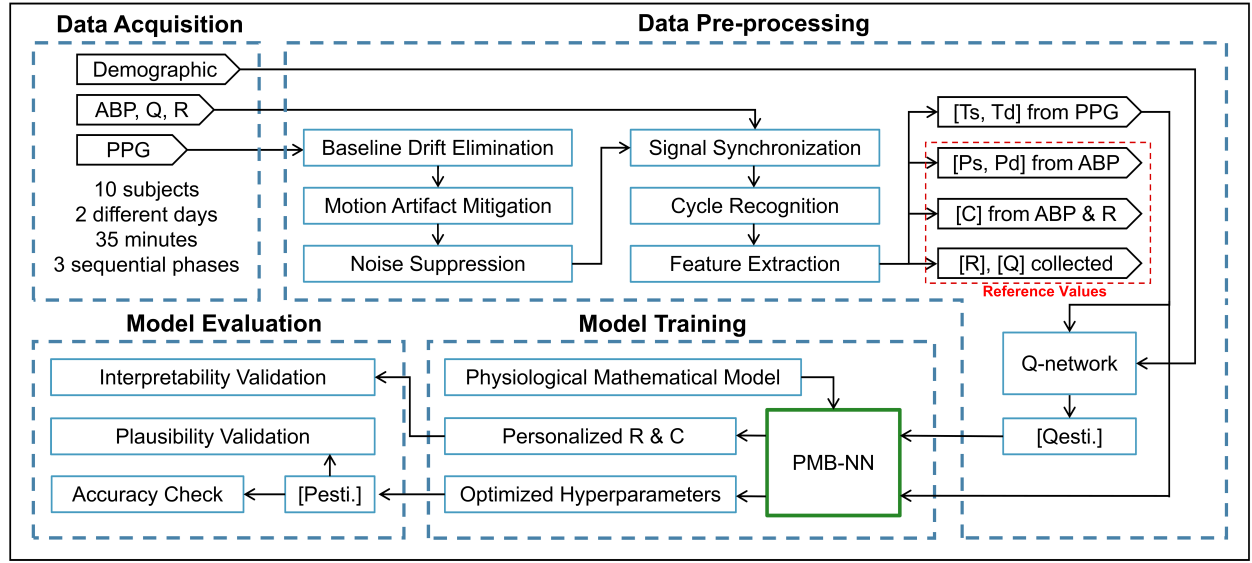

Figure 2: Schematic overview of the PMB-NN validation pipeline, delineating four core stages: (i) simultaneous acquisition of subject characteristics data (age, gender, height, weight), $Q$ and ABP, and PPG, where $Q$ and ABP are the reference values for model training and evaluation; (ii) comprehensive signal pre-processing and feature extraction to assemble model inputs: $T_s$, $T_d$, $Q_{esti.}$ (obtained from $T_s$, $T_d$ and subject characteristics through Q-network) and corresponding reference targets: $P_s$, $P_d$, $R$, $C$; (iii) PMB-NN architecture design and training; and (iv) model performance evaluation. $T_s$: systolic upstroke time; $T_d$: diastolic time; ABP: arterial blood pressure; $P_s$: systolic blood pressure; $P_d$: diastolic blood pressure; $P_{esti.}$: estimated blood pressure including systolic and diastolic; $Q$: cardiac output; $Q_{esti.}$: estimated cardiac output; $R$: total peripheral resistance; $C$: arterial compliance.

Resting brachial blood pressure as the baseline value was measured using an ambulatory BP monitor (ABPM 7100, IEM GmbH, Stolberg, Germany; distributed by Welch Allyn, USA) with the subject seated before the continuous physiological measurements. The measurements proceeded only if the Finometer-derived systolic and diastolic pressures fell within ± 20% of the baseline values from ABPM 7100, ensuring measurement validity.

## 2.2. Data pre-processing

The overall workflow of the method is illustrated in Fig. 2.

### 2.2.1. PPG pre-processing and hemodynamic reference extraction

We derived the reference values: hemodynamic parameters, including beat-to-beat systolic blood pressure ($P_s$), diastolic blood pressure ($P_d$), cardiac output ($Q$), and total peripheral resistance ($R$), from continuous arterial pressure waveform using BeatScope® Easy software (Finapres Medical Systems) incorporating the Modelflow algorithm [36]. Arterial compliance ($C$) was estimated on a beat-to-beat basis through pressure decay analysis [37] of the arterial blood pressure waveform, where $C$ equals to the decay time divided by $R$.

Concurrently, PPG signals were exported via Consensys® software. The PPG signals underwent a comprehensive pre-processing sequence to optimize signal quality: (1) baseline drift correction via linear detrending [38], (2) motion artifact suppression using a 0.5-s median filter [39], (3) spectral noise reduction through a 4th-order Butterworth bandpass filter (0.5-5.0 Hz) aligned with the Nyquist criterion [40], and (4) final smoothing via a 0.1-s moving average filter [41].

Signal synchronization was achieved by temporal alignment of device timestamps, considering physiological time delays between finger PPG and BP measurements negligible due to anatomical proximity (< 2 cm inter-sensor distance). Beat identification of PPG was performed based on BP waveform landmarks where systolic upstroke time ($T_s$) was determined as the maximum rising slope interval from systolic foot to systolic peak, and diastolic time ($T_d$) was the interval from systolic peak to next systolic foot. Accordingly, $T_d$ was calculated as the difference between beat duration ($T_c$) and $T_s$.

All derived parameters ($T_s$, $T_d$, $P_s$, $P_d$, $Q$, $R$, $C$) were processed using a 30-cycle Savitzky-Golay filter (second order polynomial) to mitigate transient fluctuations. This 30-cycle window length was selected based on two critical considerations: First, to attenuate short-term physiological noise (0.15-0.4 Hz) arising from vasomotor activity, respiratory modulation and residual motion artifacts that persisted despite experimental controls. The 30-cycle window effectively preserved hemodynamic trends while suppressing high-frequency variability unrelated to the exercise-induced cardiovascular responses [42]. Second, this approach aligns with clinical monitoring standards where 30-60 second averages are established as optimal for capturing physiologically meaningful blood pressure trends, particularly during dynamic exercise when beat-to-beat variability is amplified by locomotor cadence and metabolic demands [43].

### 2.2.2. Cardiac output estimation for PMB-NN input (Q-Network)

To obtain the cardiac output values as the input of our PMB-NN model, a fully connected neural network (FCNN), named Q-network, was developed to estimate $Q$ based on the Day-1 training dataset, including $T_s$ and $T_d$, along with subject characteristics data (age, sex, height, and weight). The Q-network was trained and tested using the Leave-One-Subject-Out (LOSO) cross validation method to both evaluate generalization performance across subjects and select the optimal model structure and hyperparameter configuration. Candidate models in Appendix A with different network depths, hidden layer sizes, dropout rates and initial learning rates were compared, using the average mean squared error (MSE) in equation 1 on the test subjects as evaluation criteria.

$$L_{1data} = \frac{1}{M} \sum_{i=1}^{M} (\hat{Q}_i - Q_i)^2 \qquad (1)$$

Where $\hat{Q}_i$ and $Q_i$ are the output and reference target cardiac output value of the $i^{th}$ sample point, respectively. $M$ is the sample size of output $\hat{Q}$. The optimal configuration was empirically selected by LOSO method, which consists of a 6-64-128-64-1 topology, a dropout probability of 0.2, and an initial learning rate of $1 \cdot 10^{-3}$. ReLU activation functions [44] were applied to the hidden layers to introduce nonlinearity and alleviate issues with vanishing gradients. AdamW optimizer [45] was applied to prevent overfitting. Learning rate was automatically adjusted by a ReduceLROnPlateau scheduler [46], reducing it by half after 20 epochs of no improvement. The training process stopped after 300 epochs or was terminated early if the error reached a predefined threshold $\epsilon = 0.1$ L/min. The Q-network achieving the lowest average MSE across all LOSO folds was selected as the final Q-network. This trained network was then applied to each individual's Day-2 dataset to obtain the estimated cardiac output ($Q_{esti.}$) for individuals.

### 2.3. Architecture of the Physiological Model-based Neural Network
### 2.3.1. Hybrid AI scheme—Physiological Model-based Neural Network (PMB-NN)

The proposed hybrid AI scheme integrates the physiological model described in 2.3.2 into the loss function of the neural network, imposing physiology-informed constraints that ensure model outputs adhere to hemodynamic principles. As illustrated in Fig. 3, the overall scheme comprises four components: inputs, training phase, testing phase and outputs.

For PMB-NN's inputs, the time feature vector $T$ was systematically generated through alternating concatenation of $T_s$ and $T_d$ in the sequence $[T_{s,1}, T_{d,1}, T_{s,2}, T_{d,2}, ..., T_{s,n}, T_{d,n}]$, where $n$ represents the total number of cardiac cycles. Similarly, the target pressure vector $P$ was constructed by interleaving $P_s$ and $P_d$ pressures in the corresponding sequence $[P_{s,1}, P_{d,1}, P_{s,2}, P_{d,2}, ..., P_{s,n}, P_{d,n}]$.

A feedforward neural network was designed to map the input signal $T$ to the estimated blood pressure $\hat{P}$. The network consists of an input layer, three hidden layers (128 units each), and an output layer (1-128-128-128-1 topology). ReLU activation functions [44] were employed for all hidden layers to introduce nonlinearity while mitigating vanishing gradients. This architecture was selected to balance model capacity and computational efficiency, as preliminary experiments showed deeper or wider layers yielded marginal improvements at higher computational costs.

To embed physiological constraints, the training objective combines a data-fitting term with two physiology-informed regularization terms derived from physiological model in . The total loss is computed as:

$$L_{tot} = L_{2data} + L_{E1} + L_{E2} \qquad (2)$$

which was minimized to jointly optimize the feedforward neural network parameters (weights and bias) and the constraints of the physiological-model parameters $R$, $C$. The data fitting loss term, $L_{2data}$, is the MSE between estimated

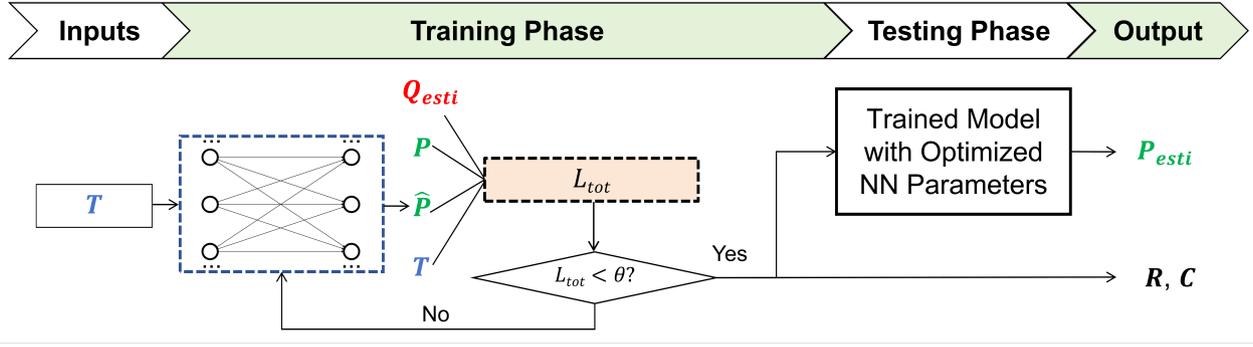

Figure 3: Diagram of the proposed PMB-NN scheme. $T$: input time feature vector. $P$: target pressure vector. $\hat{P}$: feedforward neural network output pressure vector. $Q_{esti.}$: estimated cardiac output. $P_{esti.}$: estimated blood pressure including systolic and diastolic. $R$: total peripheral resistance. $C$: arterial compliance.

$P_i$, hat and ground-truth $P_i$ ensures alignment with empirical measurements:

$$L_{2data} = \frac{1}{N}\sum_{i=1}^{N}(\hat{P}_i - P_i)^2 \quad (3)$$

where $N$ is the number of samples. The physiological regularization loss terms, $L_{E1}$ and $L_{E2}$, enforce adherence to the mathematical model of equation 9 and 10, penalizing deviations from expected hemodynamic relationships. The losses are defined as:

$$L_{E1} = \frac{1}{n}\sum_{i=1}^{n}(\hat{P}_{s,i} - f_1(P_{d,i-1}, T_{s,i}, T_{d,i}, Q_{esti,i}, R, C))^2 \quad (4)$$

$$L_{E2} = \frac{1}{n}\sum_{i=1}^{n}(\hat{P}_{d,i} - f_2(P_{s,i}, T_{d,i}, R, C))^2 \quad (5)$$

where $n$ equals to $0.5N$ from the alternating concatenation. $\hat{P}_{s,i}$ and $\hat{P}_{d,i}$ represent the estimated systolic and diastolic pressure in the $i^{th}$ cycle, respectively. $P_{d,i-1}$ and $P_{s,i}$ are ground-truth diastolic pressure in the $i-1^{th}$ cycle and systolic pressure in the $i^{th}$ cycle, respectively. $Q_{esti,i}$ is the $i^{th}$ cycle's cardiac output estimated from Q-network. $R$ and $C$ indicate total peripheral resistance and arterial compliance, respectively. The initial values of these two physiological model parameters were set as 1 based on physiological baseline of $R$ and C. Detailed parameterization procedure are provided in .

In model training, the Adam optimizer was employed with an initial learning rate of 0.01, dynamically adjusted via a scheduler to stabilize convergence. Training terminated when $L_{tot}$ fell below a threshold of 3 within 1,000 iterations, ensuring the balance between numerical precision and physiological fidelity to prevent overfitting while maintain computational efficiency. PMB-NN underwent independent train-test runs for 10 subjects regarding all 3 activity types through 30 Day-1 training sets and 30 Day-2 testing sets, individually.

### 2.3.2. Physiological mathematical model

We merged an established physiological model in previous works [47],[22] in our hybrid PMB-NN scheme. The physiological model is originated from two-element Windkessel model which encodes the RC pressure-flow relationship to represent whole body circulation dynamics [48] illustrated in Fig. 4. The model represents the relationship between the temporal features of the cardiac cycle and BP values. The relationship is mathematically described by the equation 6 and 7:

$$P_{s,i} = P(t|t = T_{s,i})$$
$$= P_{ts,i}e^{-T_{s,i}/RC} + \frac{I_{0,i}T_{s,i}C\pi R^2}{T_{s,i}^2 + C^2\pi^2R^2}\left(1 + e^{-T_{s,i}/RC}\right) \quad (6)$$

$$P_{d,i} = P(t|t = T_{d,i}) = P_{td,i}e^{-T_{d,i}/RC} \quad (7)$$

where $i$ is the number of cardiac cycle. Each cycle starts with a systolic upstroke wave while ends at the end of diastolic wave. $T_{s,i}$, $T_{d,i}$ and $T_{c,i}$ (seconds), denotes systolic upstroke time, diastolic time and cycle duration in the $i^{th}$ cycle, respectively. $P_{s,i}$ and $P_{d,i}$ (mmHg) denotes the systolic and diastolic pressure value of $i^{th}$ cycle, which locate at the peak and the termination of the $i^{th}$ cycle waveform, respectively. $P_{ts,i}$ and $P_{td,i}$ are the initial values of $P_{s,i}$ ($= P_{d,i-1}$) and $P_{d,i}$ ($= P_{s,i}$) of each cardiac cycle, respectively [22]. $R$ and $C$ are in units of mmHg·s/ml and ml/mmHg, respectively. $I_{0,i}$ is the peak value of a sinusoidal function which represents total blood pumped by heart during systole [47] in the $i^{th}$ cycle, calculated as:

$$I_{0,i} = \frac{1000Q_iT_{c,i}}{60\int_0^{T_{s,i}}\sin\left(\frac{\pi t}{T_{s,i}}\right)dt} \quad (8)$$

where $Q_i$ (L/min) is the cardiac output of $i^{th}$ heart cycle. 1000/60 is applied to convert the cardiac output unit from L/min to ml/s. Accordingly, we derived the representation for $P_{s,i}$ and $P_{d,i}$ in the general form as equation 9 and equation 10:

$$P_{s,i} = f_1(P_{d,i-1}, T_{s,i}, T_{d,i}, Q_i, R, C) \quad (9)$$

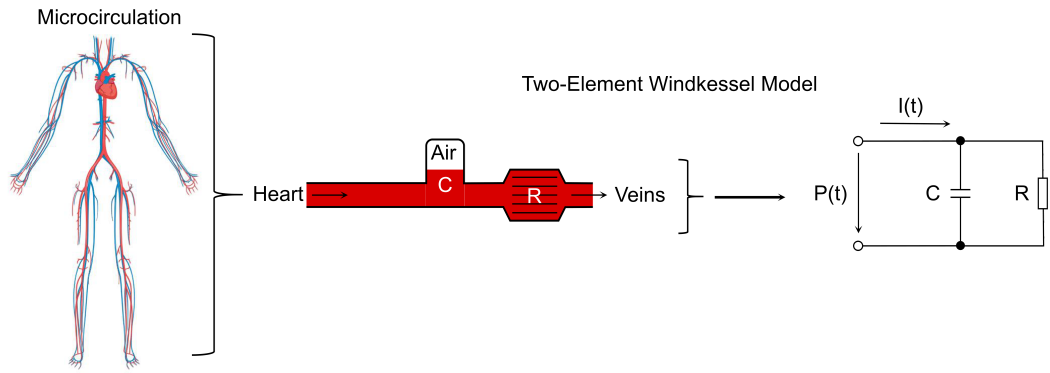

Figure 4: Schematic of the two-element Windkessel model for entire circulation dynamics.

$$P_{d,i} = f_2(P_{s,i}, T_{d,i}, R, C) \quad (10)$$

which were the representation of our physiological model (referred to as PM hereafter). PM is inherently capable of identifying key hemodynamic parameters ($R, C$) values with the given $T_s$, $T_d$, $P_s$, $P_d$ and $Q$.

### 2.4. Model evaluation

All the data pre-processing, model developing, training and evaluation were implemented on a workstation equipped with 12th Gen Intel(R) Core(TM) i7-12700H CPU. Deep learning framework Pytorch 2.1.0 in Python 3.9.13 was fully applied in establishing our algorithms.

#### 2.4.1. Benchmark models and parameterization for subsequent evaluation

To enable a multidimensional evaluation of PMB-NN, we compared it with both the standalone physiological model (PM) and representative deep learning benchmarks: FCNN, CNN-LSTM, Transformer and PM. FCNN has the same feedforward neural network structure with PMB-NN but optimize with MSE loss ($L_{2data}$) only. Transformer [49] and CNN-LSTM [50] are the commonly used structure with empirical hyperparameter decisions. All benchmark models underwent independent train-test runs through 30 Day-1 training sets and 30 Day-2 testing sets.

For the PMB-NN, the parameterization of the neural-network parameters and the physiological parameters $R$ and $C$ was performed jointly during model training. Initialized at a physiologically normalized value of 1.0, $R$ and $C$ were formally instantiated as learnable tensors (through nn.Parameter in PyTorch) within the model's computational graph, sharing the same optimization status as the synaptic weights and biases. During backpropagation, the composite loss—which includes data-fitting and physics-based terms—produced gradients with respect to both the network parameters and the physiological variables. Consequently, the Adam optimizer all parameters in a single step through a unified parameter space. This avoids any alternating or hierarchical optimization scheme.

For the benchmark DL models, (FCNN, CNN-LSTM, and Transformer) architectural and optimization hyperparameters (including layer depth, hidden units, kernel size, attention heads, learning rate described in the Appendix B) were manually specified prior to training and remained fixed according to the state-of-art [49, 50]. During training, the model parameters (weights and bias) were optimized via backpropagation to minimize the training data fitting loss. For the physiological benchmark model, PM was parameterized through the direct identification of the vascular parameters $R$ and $C$ using a curve-fitting procedure grounded in the underlying hemodynamic equations. For each data segment, optimal $R$ and $C$ values were determined via the Limited-memory Broyden–Fletcher–Goldfarb–Shanno (L-BFGS-B) optimizer by minimizing the summed squared error between measured pressures and the $P_s$ and $P_d$ iteratively simulated from $T_s$, $T_d$ and $Q$ across all cardiac cycles.

#### 2.4.2. Performance evaluation

Performance evaluation and physiological interpretability, plausibility validations were done in three categories: SP, LIC and MIC, while performance evaluation and plausibility validation of BP were systematically done across $P_s$ and $P_d$. The estimation accuracy of PMB-NN was compared against the benchmark models by evaluating the Mean Error (ME) with Standard Deviation (SD) and Mean Absolute Error (MAE) for $P_s$ and $P_d$ across each activity and subject. We assessed clinical acceptance using the AAMI standard, developed by the Association for the Advancement of Medical Instrumentation [51], which requires a mean error (ME) within ± 5 mmHg and a standard deviation (SD) below 8 mmHg to ensure accuracy comparable to manual auscultation. The percentage of samples meeting the AAMI criterion is also reported. Bland-Altman analysis was applied to quantify bias and limits of agreement for PMB-NN's estimation accuracy. P-value from the Wilcoxon signed-rank tests and Cohen's d-value were calculated in |ME|, SD and MAE's pairwise comparison between PMB-NN and other four respective models. Statistically significant difference was treated with two-tailed p-value less than 0.05, while Cohen's d values of 0.2, 0.5, and 0.8 represented small, medium, and large effects, respectively [52].

### 2.4.3. Physiological interpretability validation

To test PMB-NN's ability to identify physiologically interpretable $R$ and $C$ values across diverse activity occasions, we compared its parameter identification performance against those of sole PM's and the reference values. Estimation errors for $R$ and $C$ were calculated for each of the 30 data segments to derive the ME ± SD per category.

### 2.4.4. Physiological plausibility validation

According to the physiological plausibility, we expected to observe monotonic inverse relationships between the estimated $P_s$ and $T_s$, and between the estimated $P_d$ and $T_d$. The systolic phase during exercise is characterized by a shortened $T_s$ duration and elevated $P_s$, resulting from increased myocardial contractility that accelerates ventricular ejection. This manifests as a steeper systolic upstroke and higher systolic pressure, reflecting integrated adaptations in arterial compliance and wave-reflection dynamics [53]. Similarly, a longer $T_d$ duration permits greater peripheral run-off and thus tends to reduce $P_d$, implying an inverse association between $P_d$ and $T_d$. During dynamic exercise, metabolic vasodilation keeps or slightly increases diastolic blood pressure, while increasing heart rate shortens the diastolic interval [54].

To assess the correlation between each model's estimates, scatter plots were generated for estimated $P_s$ versus $T_s$, and estimated $P_d$ versus $T_d$. Their Spearman's correlation coefficient ($\rho$) of PMB-NN and the benchmark models were calculated. To statistically compare the correlation performance between PMB-NN and the comparative models, the Wilcoxon signed-rank test was then used to assess the differences between the paired Z-scores of the two methods. Fisher's Z-transformation was applied to the individual rho values to stabilize their variance and transform them into a metric suitable for comparative analysis. Finally, the practical magnitude of any observed statistical difference was quantified using Cohen's d for paired samples, calculated based on the difference in the Z-scores to provide an estimate of the effect size of difference. A two-sided p < 0.05 was used to determine significance which indicates a statistically difference in capturing inverse correlation between two models.

### 2.4.5. Effect of estimated cardiac output on model performance

To investigate the relationship between cardiac output estimation accuracy and the performance of the PMB-NN, MAE and mean absolute percentage error (MAPE) of $Q$ were calculated. Consistent with validation studies spanning rest to dynamic exercise, we considered MAPE ≤ 30% as clinically acceptable [55]. To evaluate the models' inter-parameter error dependency on the performance and interpretability of PMB-NN, we performed the Pearson correlation coefficient ($r$) analysis between the MAE of the estimated $Q$ and estimated BP, between the MAE of the estimated $Q$ and the absolute error of $R$, and between the MAE of the estimated $Q$ and the absolute error of C.

## 3. Results

### 3.1. Subject characteristics

The baseline characteristics of the subjects, including sex, age, height, as well as the interval between two measurements, and weight, the left upper arm (LUA) $P_s$ and $P_d$ from both the first (1st) and the second measurement (2nd), are summarized using median, maximum and minimum values in Table 1.

| Sex | 7 male, 3 female | | |
|---|---|---|---|
| Characteristic | Median | Max. | Min. |
| Days from 1st to 2nd | 110 | 126 | 20 |
| Age (years) | 25 | 29 | 20 |
| Height (cm) | 173.8 | 193.5 | 159.0 |
| Weight 1st (kg) | 68.6 | 93.9 | 60.0 |
| Weight 2nd (kg) | 70.8 | 95.0 | 62.3 |
| LUA $P_s$ 1st (mmHg) | 120 | 126 | 109 |
| LUA $P_s$ 2nd (mmHg) | 78 | 85 | 65 |
| LUA $P_d$ 1st (mmHg) | 121 | 128 | 109 |
| LUA $P_d$ 2nd (mmHg) | 76 | 82 | 69 |

Table 1: Summary of subject and baseline characteristics

Fig. 5 illustrates the distribution of the $P_s$ and $P_d$ values across different activity types in the entire database. The analysis comprises 19483, 20109, and 23674 data points (BP cycles) for the static posture (SP), low intensity cycling (LIC), and moderate intensity cycling (MIC) conditions, respectively. For $P_s$ in Fig. 5a, the mean, Q1, Q3 are 111.9, 104.0, 117.6 mmHg for SP, 136.6, 128.9, 144.2 mmHg for LIC, 150.0, 137.8, 162.2 mmHg for MIC. For $P_d$ in Fig. 5b, the mean, Q1, Q3 are 71.6, 67.2, 76.6 mmHg for SP, 73.4, 68.7, 78.5 mmHg for LIC, 78.6, 71.6, 85.1 mmHg for MIC.

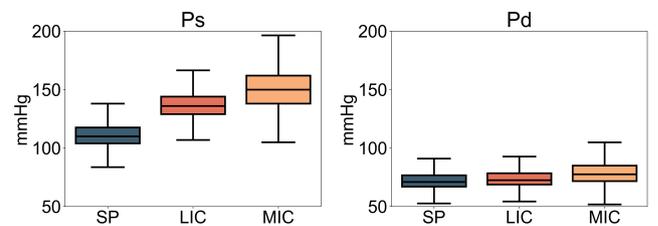

Figure 5: Distribution of $P_s$ (left) and $P_d$ (right) for different activity types from the whole database.

### 3.2. Accuracy evaluation

The detailed error metrics for PMB-NN, FCNN, CNN-LSTM, Transformer and PM, for each subject and activity, with their median, minimum and maximum values are listed in Table C.1, Appendix C. Overall for PMB-NN, median (min, max) MAE are 7.2 (4.1, 15.0) mmHg for $P_s$ and 3.9 (1.9, 6.4) mmHg for $P_d$. 39.32% $P_s$ estimation samples and 69.09% $P_d$ estimation samples fell within AAMI standard (within ± 5 mmHg). 23/30 of the $P_s$ estimation segments and 30/30 of the $P_d$ estimation meet the AAMI criterion.

It was also observed that PMB-NN's estimation accuracy decreases with exercise intensity increase particularly for $P_s$ given the increase of BP amplitude, meanwhile $P_s$ estimations show greater variability than $P_d$. Meanwhile, PMB-NN, FCNN, CNN-LSTM and Transformer maintained most ME values within established AAMI threshold limits, while the PM model exhibited systematically higher ME values, indicating reduced estimation accuracy. The SD distribution for PMB-NN, FCNN and CNN-LSTM's $P_d$ estimations fell entirely within AAMI-recommended thresholds, demonstrating exceptional measurement precision.

The comparative performance of PMB-NN and the benchmark models are shown in Fig. 6. Table C.1 in Appendix C listed all the five models' accuracy metric values, comparison in general and respect to each activity type. Statistical comparison of |ME|, SD and MAE values in Fig. 6 yielded: For $P_s$ estimation, PMB-NN significantly outperformed the PM across all activities. PMB-NN revealed significant lower SD and MAE comparing with Transformer in SP and LIC, while non-significant differences of all activities versus FCNN and CNN-LSTM. For $P_d$ estimation, PMB-NN showed non-significant differences versus Transformer across all activities, significantly outperform PM in MIC, while significantly underperform FCNN in SP, MIC, and CNN-LSTM in LIC, MIC.

Fig. 7 shows the average of all the 30 testing sets for both measured pressure and all models' estimated pressure. Across all activity conditions, both $P_s$ and $P_d$ estimated by PMB-NN, FCNN and CNN-LSTM converge toward a stable central tendency within each task. Owing to the lower intrinsic variability of $P_d$, their estimates exhibit high accuracy. On the other hand, $P_s$ estimation shows greater fluctuation particularly under exercise, yielding marginally reduced accuracy relative to $P_d$. Nevertheless, their reliably captures the systematic shift in blood pressure ranges from SP to LIC to MIC, reflecting the expected stepwise increase with rising exercise intensity. However, Transformer exhibits greater variability in $P_s$ estimates, although it better matches the reference $P_s$ during the first half of LIC and MIC. Meanwhile for PM, the estimation curves exhibit noticeable irregular oscillations. This effect is particularly pronounced under MIC, where the PM estimates deviate more substantially from the measured values compared with PMB-NN and other three benchmarks.

The Bland-Altman analysis in Fig. 8 reveals distinct characteristics in the PMB-NN's blood pressure estimation performance. For $P_s$, the model demonstrated a mean bias of 0.1 mmHg and a wider limits of agreement (LoA) within +20.6 mmHg and -21.1 mmHg, indicating no significant systematic overestimation or underestimation and large variability in certain samples. For $P_d$, the model demonstrated a mean bias of -0.3 mmHg and narrower LoA within +11.5 mmHg and -10.6 mmHg. The error distribution was relatively stable across the measurement range.

Further insights were revealed by the detailed Bland-Altman analysis for 10 subjects, 3 activities. $P_d$ estimates show tight agreement with the references across all activities: point clouds are compact, mean bias is close to zero, and the dispersion remains relatively constant from SP to MIC, indicating little or no proportional bias. In contrast, $P_s$ exhibits wider dispersion, particularly during LIC/MIC, consistent with higher physiological variability at increased workloads. All subjects display a positive slope of error versus mean for $P_s$, suggesting mild proportional bias attributable to the relatively homogeneous structure of the input signals and the larger fluctuation range of the measured blood pressure values. Taken together, the BA plots indicate high accuracy and stability for $P_d$, and good but comparatively less tight agreement for $P_s$.

### 3.3. Physiological interpretability validation

The comparative analysis demonstrates distinct performance characteristics between PMB-NN and PM in estimating $R$ and $C$ across static and dynamic conditions (Fig. 9). PMB-NN exhibits high estimation accuracy for $R$, maintaining mean errors within 0.15 mmHg·s/ml under LIC and MIC condition (0.10 ± 0.21 mmHg·s/ml for LIC and 0.11 ± 0.24 mmHg·s/ml for MIC) and showing largest error escalation during SP (-0.17 ± 0.39 mmHg·s/ml). PM exhibits similar accuracy, with error of -0.15 ± 0.20 mmHg·s/ml, 0.11 ± 0.12 mmHg·s/ml and 0.12 ± 0.13 mmHg·s/ml for SP, LIC and MIC, respectively. Bland-Altman analysis further underscores the precision of both methods, with PMB-NN and PM exhibiting similarly narrow 95% limits of agreement across all test conditions, indicative of high fidelity in $R$ estimation.

In contrast, $C$ estimation proves more challenging, where both models systematically underestimates C. For PMB-NN, the errors are -0.12 ± 0.27 ml/mmHg, -0.46 ± 0.21 ml/mmHg and -0.47 ± 0.25 ml/mmHg for SP, LIC and MIC. For PM, the errors are -0.10 ± 0.28 ml/mmHg, -0.46 ± 0.20 ml/mmHg and -0.45 ± 0.24 ml/mmHg for SP, LIC and MIC. Both PMB-NN and the conventional PM model display a nearly identical underestimation profile in LIC and MIC. Notably, both architectures accurately track the anticipated decline in $R$ and $C$ values as exercise intensity increased.

### 3.4. Physiological plausibility validation

Statistical comparison reveals that PMB-NN significantly better captured $P_s$-$T_s$ negative correlation comparing with CNN-LSTM (p = 0.018, d = -0.463), Transformer (p < 0.001, d = -69.335) and PM (p < 0.001, d = -27.230) while comparable with FCNN (p = 0.317, d = -0.183). Meanwhile, PMB-NN significantly better captured $P_d$-$T_d$ negative correlation comparing with all the benchmark models: FCNN (p = 0.015, d = -0.477), CNN-LSTM (p < 0.001, d = -0.890), Transformer (p < 0.001, d = -5.166) and PM (< 0.001, d = -5.201). As shown by the scatterplots of Fig. 10 and corroborated by the per-segment Spearman coefficients in Table D.1, Appendix D, PMB-NN yields the anticipated exercise-physiology pattern that most consistent with $P_s$-$T_s$ and $P_d$-$T_d$ relations showing clear downward trends and mean Spearman's $\rho$ = -1 for $P_s$-$T_s$ relation and -0.971 for $P_d$-$T_d$ relation. FCNN similarly captures the expected negative $P_s$-$T_s$ monotonicity ($\rho$ = -0.98), but its $P_d$-$T_d$ relation ($\rho$

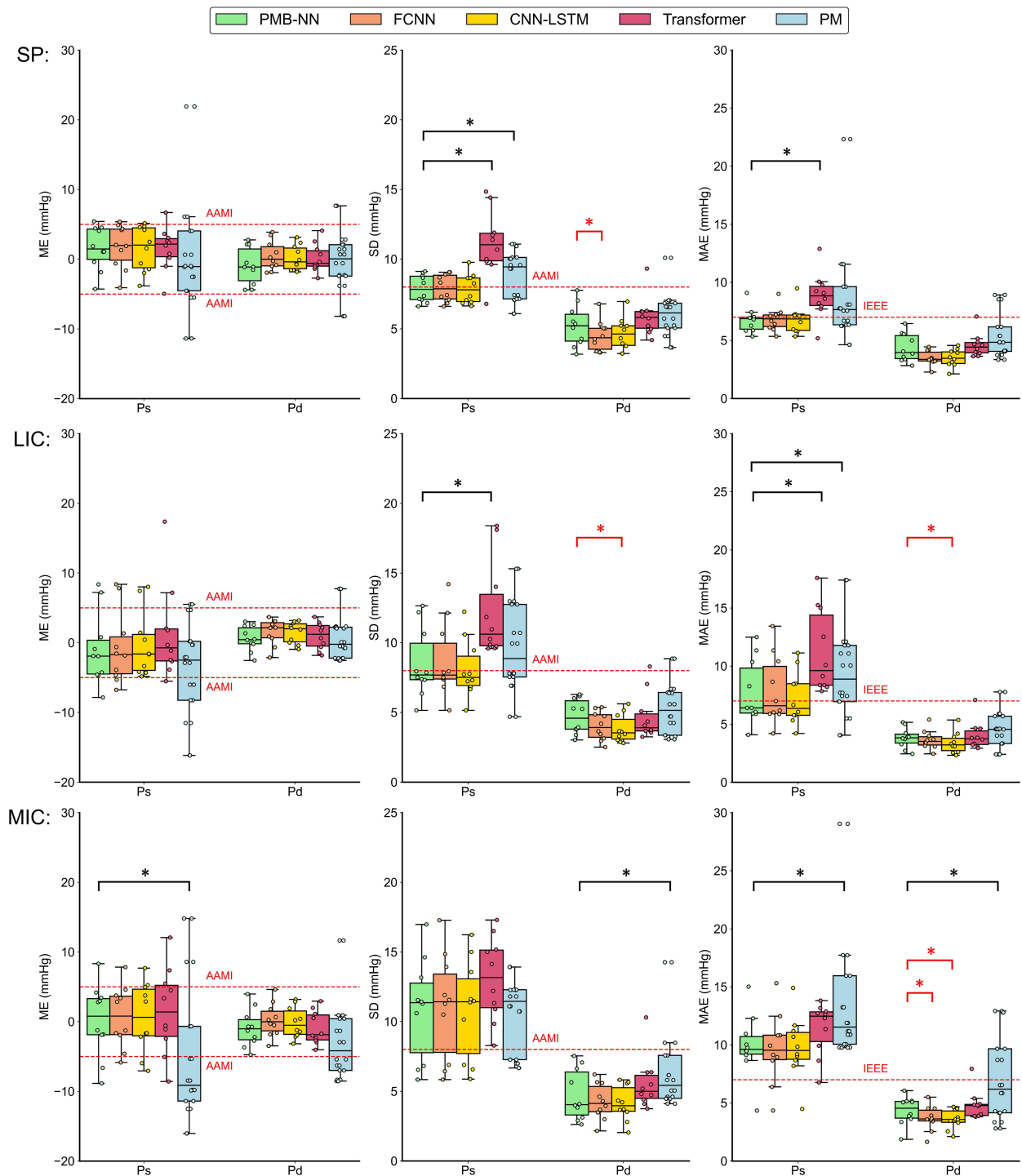

Figure 6: Comparative performance evaluation between PMB-NN and FCNN, CNN-LSTM, Transformer, PM models for each activity (SP = static posture, LIC = low-intensity cycling, MIC = moderate-intensity cycling) across ME, SD, and MAE metrics. 10 points in each box from each of the nine subfigures represent the results' metrics on 10 subjects' testing sets. From the Wilcoxon signed-rank test for comparison between PMB-NN and the benchmark models, significant difference (p < 0.05) is shown using (∗). Black and red (∗) indicate PMB-NN's significantly outperforming and underperforming, respectively. Absolute values for ME are denoted in the statistical comparison. The AAMI and IEEE cuffless wearable standard limits are shown where applicable.

= -0.548) is relatively unstable, even with 5/30 segments reversing to positive $\rho$, contrary to expectation. Similar with FCNN, CNN-LSTM captured better $P_s$-$T_s$ correlation ($\rho$ = -0.86) comparing with $P_d$-$T_d$ correlation ($\rho$ = -0.451). The

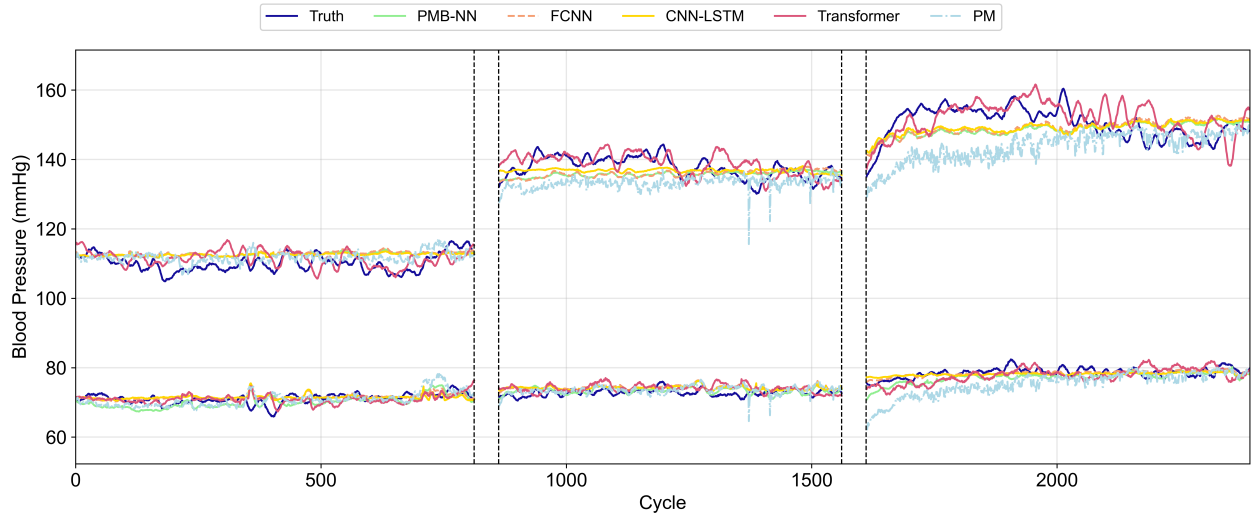

Figure 7: Aggregated $P_s$ (upper) and $P_d$ (lower) performance from all models' estimation comparing with measured data across 30 segments (10 subjects × 3 activity types). The x-axis is cardiac cycle index. Three blocks in each row correspond to SP (Static Posture), LIC (Low-Intensity Cycling), and MIC (Moderate-Intensity Cycling).

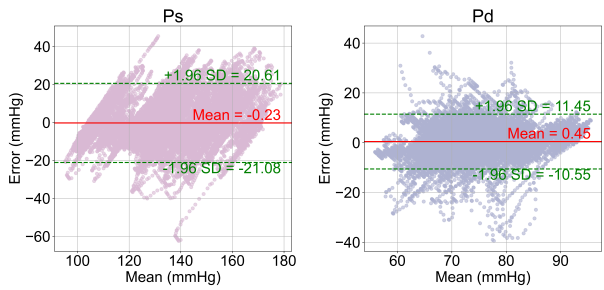

Figure 8: Performance evaluation of the PMB-NN model for blood pressure estimation. Bland-Altman plots reveal the agreement between estimated and reference values for $P_s$ (left) and $P_d$ (right). Dashed lines represent the 95% limits of agreement (± 1.96 SD), with solid lines indicating mean bias.

Transformer exhibits weak and highly variable structure for both $P_s$-$T_s$ ($\rho$ = -0.009) and $P_d$-$T_d$ ($\rho$ = -0.160), with $\rho$ values scattered in sign and magnitude, consistent with the diffuse, unpatterned point clouds. Finally, the PM model shows largely random $P_s$-$T_s$ correlations ($\rho$ = -0.208) yet captures the diastolic relation better ($\rho$ = -0.778), producing predominantly negative $P_d$-$T_d$ correlations that align with physiological intuition.

### 3.5. Cardiac output accuracy evaluation

Fig. 11 represents the cardiac output estimation accuracy from the Q-network with the optimal configuration. Median (min, max) values for MAE is 1.12 (0.37, 3.26) L/min and for MAPE is 12.79 (5.19, 27.33) %. The results reveals subject variability. Large errors between estimated and reference $Q$ occurs in each activity type, e.g., subject 1 in LIC/MIC, subject 3 and 10 in SP. Nevertheless, all MAPE values remained ≤ 30%, indicating errors within commonly accepted clinical limits. When stratified by activity, SP showed the highest MAPE (median 15.82%), whereas LIC and MIC yielded comparable median MAPEs (11.40% and 11.15%, respectively).

The correlation results for the MAE of $Q$ with PMB-NN estimation errors are given as follows: the correlation with $P_s$ MAE was $r$ = 0.09, with $P_d$ MAE $r$ = -0.17, with $R$ absolute error $r$ = 0.54, and with $C$ absolute error $r$ = 0.31. These results suggest that PMB-NN's blood pressure estimations are largely insensitive to $Q$ estimation errors, while the accuracy of $R$ and $C$ exhibits a moderate positive association with $Q$ estimation accuracy.

## 4. Discussion

### 4.1. Model evaluation synthesis and clinical relevance

Our findings demonstrate that the PMB-NN achieves systolic BP accuracy comparable to deep learning (DL) benchmark models, while showing lower accuracy for diastolic BP. Importantly, despite this reduced diastolic performance, PMB-NN consistently exhibited substantially higher physiological plausibility than all DL baselines, suggesting that the superior numerical accuracy of DL approaches may largely reflect overfitting to noisy signal patterns rather than physiologically coherent estimation. Beyond BP estimation, PMB-NN preserved mechanistic interpretability by identifying $R$ and $C$ with accuracy similar to the standalone physiological model (PM), while also compensating for the PM's difficulty in capturing nonlinear waveform–pressure dynamics introduced by model simplifications. Notably, PMB-NN even surpassed the PM in physiological plausibility, indicating that the hybrid architecture effectively integrates and extends the strengths of both physiological and data-driven modelling. Together, these results position PMB-NN as a balanced and physiologically grounded alternative to purely data-driven approaches for daily-life hemodynamic monitoring.

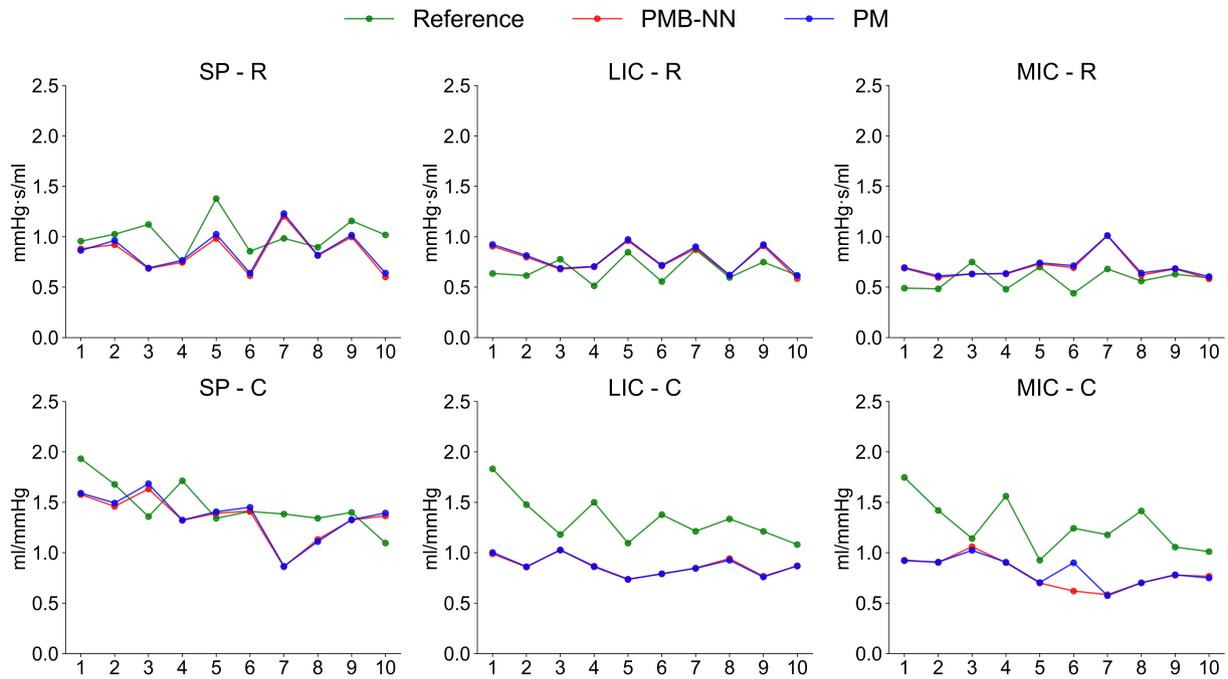

Figure 9: Estimated and reference values of *R* and *C* for each subject under different activities.

Notably, the PMB-NN does not combine its physiological and neural components uniformly across all conditions; instead, it appears to leverage the strengths of its components asymmetrically: for the artifact-sensitive $P_s$, it predominately leverages the FCNN's feature learning to bypass the systematic disturbances of the standalone PM, whereas for $P_d$, it harnesses the PM's Windkessel constraints to enforce hemodynamic validity. PMB-NN contributes to the nonlinearity or the residual estimation that the Windkessel model could not capture and meanwhile improve the overall performance of PM. This asymmetric utilization suggests that the hybrid design is not merely a linear blend of physiological and neural components, but a context-adaptive mechanism that selectively engages each component according to its physiologic relevance.

The core innovation of the PMB-NN lies in its learning direction, governed by composite loss functions that penalize deviations from the pressure-decay relationship. This effectively anchors the neural network outputs to biophysical realities. In pure data-driven models, the loss function minimizes statistical error, which can lead to "shortcut learning" where the model associates motion artifacts with pressure changes. By embedding the Windkessel constraints into the loss landscape, the PMB-NN forces the feedforward neural network to learn features that are consistent with hemodynamic decay, effectively acting as a regularizer. This explains why the PMB-NN mitigated the systematic underestimation of $P_s$ observed in the standalone PM while preventing the non-physiological volatility observed in the Transformer. This approach represents a principled trade-off: embedding domain knowledge sacrifices the model's ability to fit training data perfectly (including noise) to ensure that the generalized estimations remain within the bounds of human physiology.

An additional strength of our work lies in the experimental design. The data were collected from each subject on two separate days, with intervals ranging from 20 days to 126 days, introducing natural day-to-day physiological variability that is typically absent in controlled laboratory studies. This configuration closely mimics actual rehabilitation and health monitoring settings, where measurements are frequently made over long periods of time and across several sessions. This cross-day structure provides a realistic assessment of model robustness, and offers a strong foundation for evaluating the suitability of our models for longitudinal and daily-life monitoring. Future investigations will examine the impact of the time interval between sessions on blood pressure estimation accuracy, aiming to determine the optimal schedule for algorithm recalibration.

From a clinical perspective, reliable continuous health monitoring requires not only numerical accuracy but also plausibility, and physiological transparency to track hemodynamic changes. The improved physiological coherence and cross-day robustness of PMB-NN are particularly relevant for real-world continuous blood pressure and hemodynamic monitoring. Wearable PPG-based BP estimation often suffers from variability, limited generalization, and context-dependence, making physiological plausibility more critical than short-term numerical accuracy. By embedding physiological constraints while allowing data-driven correction, PMB-NN provides more trustworthy trajectories for trend tracking, early vascular dysfunction detection, and personalized hypertension management in ambulatory settings. Although PMB-NN only identify $R$ and $C$ estimates during the model training to embed hemodynamic constraints that help the model generate BP estimates, in future work, the PMB-NN scheme can show its dual estimation capability, which simultaneously estimate BP, $R$, and $C$, through extending PMB-NN from an offline training paradigm to an online learning framework. The model could gradually update subject-specific $R$ and $C$ during daily-life monitoring, transforming these parameters from training-only constraints into dynamic hemodynamic biomarkers. This capability would address a critical gap in wearable technologies which traditionally ignore hemodynamic parameters despite their prognostic value and enable longitudinal tracking of vascular resistance and compliance in response to ageing, lifestyle modifications, or therapeutic interventions, thereby further strengthening the potential of hybrid physiological–AI models for personalized cardiovascular care.

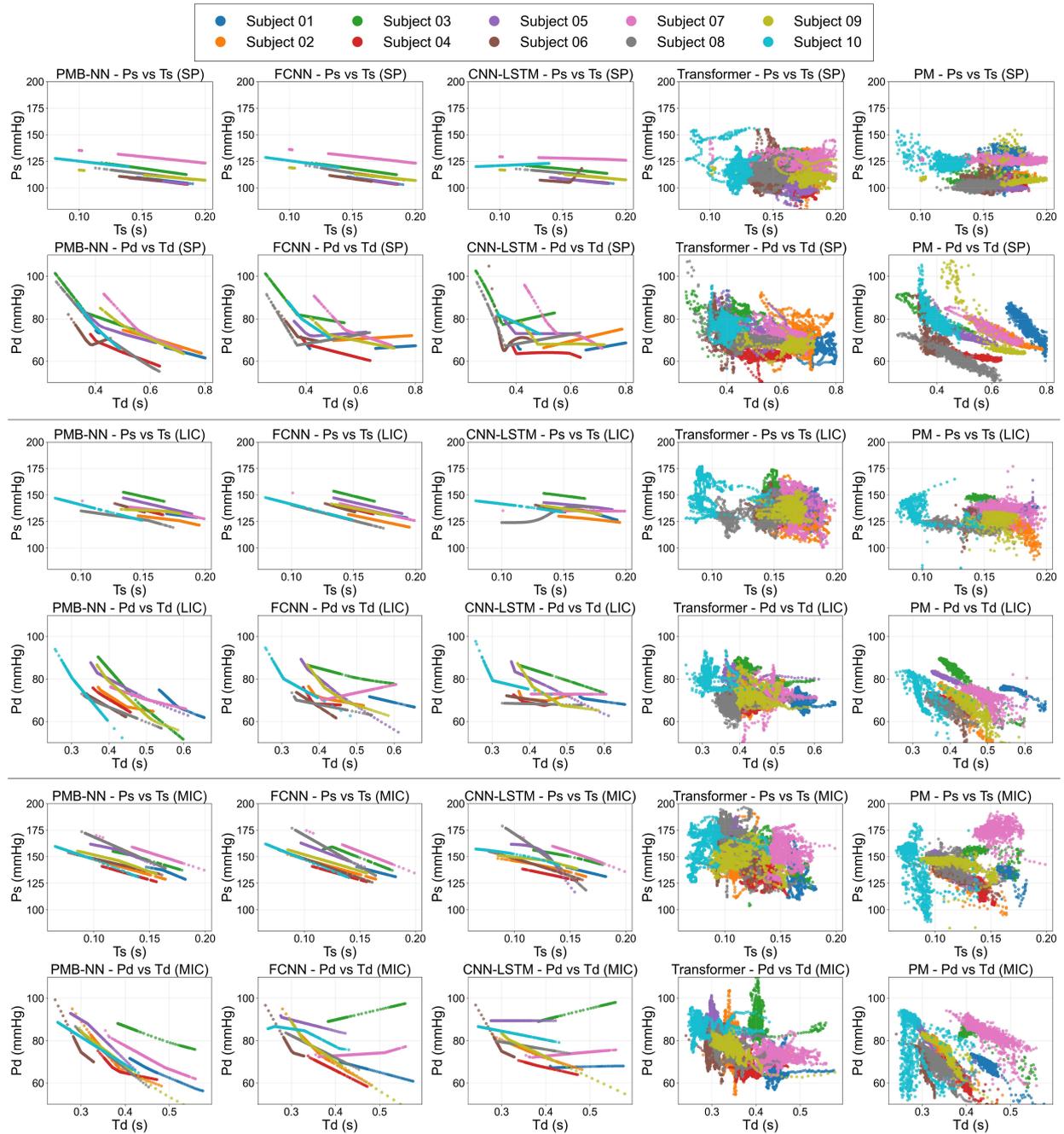

Figure 10: Scatter plots of each model's estimated pressure versus input timing. The top, middle and bottom two rows shows $P_s$ versus $T_s$ and $P_d$ versus $T_d$ for SP, LIC and MIC respectively. From left to right, the columns correspond to PMB-NN, FCNN, CNN-LSTM, Transformer, and PM, respectively. Each plot displays 10 segments from 10 subjects, distinguished by different marker colors.

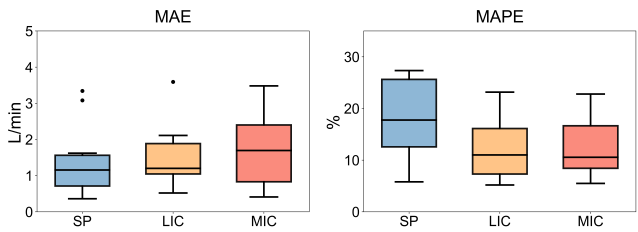

Figure 11: MAE and MAPE of $Q$ estimation from the Q-network respect to different activities for each subject, along with the summarized distribution. Subject indices are shown on the x-axis, with blue, red, and yellow points indicating SP, LIC, and MIC conditions, respectively.

### 4.2. Deep learning architectural choices: the role of the neural network

The Hybrid AI framework is designed to combine the inductive bias of physics with the learning capacity of neural networks. Our comparative analysis reveals that for low-dimensional, low-entropy data (such as derived PPG features), simpler architectures could outperform complex ones. Compared with the deep learning benchmarks, PMB-NN achieved $P_s$ performance comparable to the FCNN and CNN-LSTM, while generally superior to the Transformer. The FCNN demonstrated inherent advantages in direct pointwise regression, while the CNN-LSTM effectively utilized shallow temporal convolutions as local trend detectors. Conversely, the Transformer proved over-parameterized for this specific task, exhibiting a distinct trade-off between responsiveness and stability. While its global attention mechanism enabled rapid tracking of non-stationary transitions during exercise onset, the lack of recurrent inductive bias (inherent to LSTMs) rendered the model hypersensitive to input fluctuations. This absence of temporal 'inertia' caused the model to amplify cycle-to-cycle jitter into oscillatory noise, resulting in greater variability in $P_s$ estimates despite its capacity to capture global trends. On the other hand, the robust performance of the shallower PMB-NN and FCNN underscores the principle of model parsimony. For low-dimensional inputs, the vast parameter space of deep architectures becomes a liability rather than an asset, creating a sparse optimization landscape where the model struggles to distinguish between meaningful physiological dynamics and stochastic noise. These findings indicate that heavier AI models may not intrinsically behave better for physiological signal processing unless the input data is sufficiently rich (e.g., raw spectro-temporal embeddings).

### 4.3. Mechanistic perspective on physiological model behavior

The choice of the physiological model determines the upper bound of the system's interpretability. We employed a two-element Windkessel model primarily due to its simplicity, which aligns perfectly with the low-dimensional input features of the PMB-NN hybrid architecture, allowing for a seamless integration of physiological priors. While the PMB-NN and PM successfully tracked accurate individualized $R$, the estimation of $C$ was less accurate. While PM captures some degree of negative correlation, there are some noticeable unreasonable estimations. For example, in the 10-MIC estimation, PM estimated a wide range of $P_d$ values (from 52 mmHg to 95 mmHg) within a very small range of $T_d$ (0.25 s to 0.3 s), which clearly contradicts physiological expectations. This inconsistency likely stems from error propagation inherent in the iterative computation. Large inaccuracies in $R$ and $C$ are amplified across cycles due to the recursive dependence on previous pressure states, leading to divergent estimations. These limitations highlights the theoretical deficiency of the 0-D Windkessel model compared to 1-D tube-load or transmission line models. While finger arterial pressure provides a validated research proxy for brachial pressure, peripheral waveforms exhibit systematic deviations from central aortic pressures due to wave reflection phenomena and distal amplification [56]. This discrepancy also introduces uncertainty in $C$ estimation, as $C$ physiologically depends on central pressure-volume relationships rather than peripheral pulsatility. Consequently, exercise-induced $C$ reductions observed in our study may partially reflect peripheral waveform distortions rather than true aortic stiffening [57]. To improve interpretability, future work must move beyond the basic Windkessel model by incorporating physiologically superior paradigms, such as adaptive transfer functions for central pressure estimation or 1-D transmission line models. These refinements are essential to mitigate the impact of distal amplification and provide a more faithful representation of central hemodynamics.

Furthermore, the inadequate estimation of $Q$ compromised the accuracy of $R$ and C. This error propagation stemmed from the limitations in deriving $Q$ from PPG features and subject characteristics. Notably, the poor estimation accuracy of $Q$ was directly mirrored in the results for C. For instance, in LIC and MIC for subject 1 and 4, and in SP for subject 3. Our analyses indicated that inaccuracies in $Q$ propagate asymmetrically. $C$ exhibits high sensitivity owing to its inverse relationship with $Q$ in the Windkessel equations, where errors amplify during dynamic states like exercise-induced tachycardia. To mitigate this error propagation, future work will focus on enhancing $Q$ estimation accuracy by exploiting multi-level PPG representations and attention-based architectures to capture more robust hemodynamic patterns. Furthermore, we aim to improve model generalizability by training on larger and more diverse cohorts to better encompass physiological variability.

By tracking exercise-induced reductions in $R$, PMB-NN captures dynamic cardiovascular adaptations consistent with vasodilation during physical exertion [7]. This capability is particularly relevant for hypertensive patients, where exaggerated $R$ responses to exercise correlate with elevated cardiovascular risk [7]. Similar trends of $C$ reduction were observed in both reference data and PMB-NN's estimates, however, the literature providing a clear description of C's response to exercise remains limited. From a medical perspective, arterial compliance is primarily determined by the structural composition of the aortic wall (e.g., collagen-elastin ratio) [58]. During acute exercise, however, increased sympathetic tone enhances vascular smooth muscle contraction in the aortic wall, leading to transient stiffening and reduced compliance [59], which is a mechanism analogous to that observed in hypertensive states. Additionally, vascular mechanical properties exhibit time-dependent hysteresis, suggesting that heart rate acceleration during exercise may further modulate compliance independently of structural or neural factors [60]. Collectively, the exercise-induced reduction in $R$ and $C$ represents an adaptive hemodynamic response that optimizes cardiovascular performance during physical exertion.

### 4.4. Cross-system consistency of PMB-NN behaviour

It is noteworthy that the behaviour of PMB-NN observed in this study echoes patterns reported in our prior $\dot{V}O_2$ driven HR estimation work [32]. In both physiological subsystems, PMB-NN consistently achieved BP/HR estimation performance comparable to benchmark deep learning models while outperforming the standalone mechanistic model. More importantly, across both tasks, PMB-NN demonstrated substantially higher physiological plausibility than purely data-driven approaches, reflected by more coherent latent parameter dynamics and better adherence to expected input–output physiological relationships. These consistent cross-system observations suggest that physiology-centered hybrid AI modelling may possess stable behavioural characteristics that generalize beyond a single physiological domain. Future work will systematically investigate the internal mechanism of PMB-NN by comparing its behaviour across multiple types of physiological models and examining how differences in mechanistic structure influence performance, interpretability, and physiological plausibility. Such analyses will help reveal how physiological constraints shape the learned neural representations, where the neural component compensates for limitations in the mechanistic model, and under what conditions hybrid modelling offers advantages beyond purely data-driven approaches.

### 4.5. Limitations and future directions

While the PMB-NN demonstrates the potential of hybrid architectures, several limitations must be addressed for clinical deployment:

#### 4.5.1. Consistent bias in systolic pressure estimates

Bland–Altman analyses revealed a consistent, proportional bias in $P_s$ estimates from the PMB-NN across subjects and activities, which is an overestimation at lower means and an increasing underestimation at higher means. This suggested that the model captures the monotonic $P_s$–$T_s$ relation but remains imperfectly calibrated in scale and offset, especially at high workloads. Likely contributors include (i) limited coverage of extreme $P_s$ ranges in training, forcing extrapolation during vigorous exercise; (ii) residual nonlinearity not represented by the current mapping from timing features to pressure under changing arterial compliance and wave-reflection timing; and (iii) small but systematic time misalignment between reference and estimates during rapid transitions, which can mimic proportional errors in BA space. In addition, inter-individual differences in vascular mechanics may require subject-specific calibration, and reference $P_s$ itself may carry noise that propagates into bias. In the future, we will give priority to refining the PMB-NN's calibration by implementing piecewise or monotonic non-linear mapping layers and intensity-aware weighting directly into the model's output stage to correct the proportional bias. Furthermore, we intend to encode known physiological priors such as HR-dependent constraints and wave-reflection indices, into the network's loss function and feature space to constrain predictions during vigorous exercise.

#### 4.5.2. Critical blood pressure values

Continuous finger blood pressure measurement are widely applied for both clinical and research usage since invasive 'golden standard' measurement [61] is unsuitable for use in routine screening of large populations or clinical diagnosis. Availability of devices for continuous BP measurement from Finapres company were validated [62]. A small portion of the finger BP data collected in our experiments fell at the physiological boundary of normal ranges. This was especially noticeable during prolonged exercise sessions, where measurements were influenced by inevitable motion artifacts from upper-body movements. Additionally, extended cuff compression on the finger may cause local ischemia and cooling, which could compromise the representativeness of finger BP relative to ABP. Alternative measurement sites or wearable designs could be considered to reduce the impact of prolonged compression while maintaining signal fidelity. Moreover, experimental protocols could be adjusted by temporarily releasing the finger cuff after a period of continuous collection to allow tissue recovery, or by alternating measurement locations to minimize local ischemia.

#### 4.5.3. Restricted experimental scenarios

The requirement for continuous finger arterial pressure waveform recording required strict upper-body immobilization (particularly of the left arm) to minimize hydrostatic artifacts and motion noise, thereby limiting exercise testing to low- and moderate-intensity cycling. This constraint hindered validation during high-intensity activities or dynamic daily movements, where hemodynamic responses exhibit non-linearities essential for assessing vascular adaptability. Consequently, the model's performance for $R$ and $C$ estimation under more daily living scenarios remains unverified, potentially impacting its generalizability to real-world scenarios involving vigorous motion. In order to address this, it is considered to use accelerometer-gyroscope fusion to implement adaptive motion artifact suppression algorithms. Validation of the proposed model on miniature wrist-worn PPG devices also enables evaluation of its resistance to positional sensitivity.

*4.5.4. Sample subject characteristics and generalizability*

The model's validation cohort (young, healthy adults between 20 and 29 years old) limits clinical applicability to hypertensive and elderly populations where arterial stiffening and signal attenuation alter hemodynamics. Age-related atherosclerosis degrades PPG signal quality [60], while comorbidities like diabetes or chronic kidney disease introduce morphological aberrations that are not captured in healthy physiology [63]. These factors may challenge the Windkessel model's quasi-steady assumptions during rapid pressure fluctuations. To bridge this gap, future validation must prioritize diverse cohorts with cardiovascular pathologies and aging phenotypes. Transfer learning from clinical datasets coupled with domain-adaptive regularization will be essential for real-world application.

*4.5.5. Signal processing constraints*

The fixed 30-beat Savitzky-Golay filter, while effective for suppressing high-frequency noise in steady-state conditions, intrinsically attenuates rapid hemodynamic transitions during dynamic exercise onset, for example, abrupt sympathetic activation in cycling tests. Such smoothing may obscure essential physiological events, potentially underestimating true fluctuation magnitudes in vascular responses. This limitation arises because the filter's static window size cannot resolve time-varying signal dynamics correlated with heart rate acceleration or motion artifacts. To preserve transient fidelity without sacrificing noise robustness, adaptive filtering strategies like heart rate-tuned Kalman filters [64] can dynamically adjust smoothing windows based on R-R interval volatility. Furthermore, particle filters incorporating inertial sensor data can distinguish motion artifacts from physiological waveforms [65].

## 5. Conclusion

The PMB-NN scheme represents a convergence of physiological modeling and deep learning, offering a robust and interpretable solution for wearable hemodynamic monitoring. PMB-NN achieves systolic accuracy comparable to deep learning benchmarks while demonstrating superior physiological plausibility. Meanwhile, PMB-NN preserves mechanistic interpretability by identifying vascular parameters, effectively compensating for the physiological model's inability to capture nonlinear dynamics. PMB-NN offers a balanced, physiologically grounded alternative to black-box approaches. Future work will prioritize advanced calibration strategies to further enhance robust daily-life hemodynamic monitoring.

## A. Candidate architectures for Q-network

Specifically, the following candidate architectures and hyperparameters for Q-network in Table A.1 were considered:

| Hidden layers | Units per layer | Dropout rate | Initial learning rate |
|---|---|---|---|
| 3 | [64, 128, 64] | 0.2 | 1e-3 |
| 3 | [128, 128, 128] | 0.2 | 1e-3 |
| 3 | [128, 256, 128] | 0.3 | 5e-4 |
| 3 | [32, 64, 32] | 0.1 | 1e-3 |
| 2 | [64, 64] | 0.3 | 1e-4 |

Table A.1: List of candidate architectures and hyperparameters for Q-network

## B. Architecture of benchmark deep learning models

The FCNN model maps input T through 3 successive fully-connected hidden layers (128 units each) with ReLU activation to capture non-linear relationships, followed by a final single-unit linear output layer, trained with the Adam optimizer (lr=0.01) and a StepLR scheduler for 1000 epochs using MSE loss.

The CNN-LSTM model processes input T through a two-layer convolutional neural network (64 and 32 output channels with kernel size 3) for local feature extraction, followed by a single-layer LSTM with 256 hidden units to capture temporal dependencies, before final estimation through two fully-connected layers (128 and 1 units), trained with Adam optimizer (lr=0.01) for 1000 epochs using MSE loss.

The Transformer model processes input T through a linear embedding layer that maps the input size of 1 to a hidden dimension of 32, followed by sinusoidal positional encoding to incorporate positional information. The core architecture consists of 2 stacked Transformer encoder layers with 2 attention heads each, using self-attention to learn global temporal dependencies. The hidden representation corresponding to the input timestep is projected via a linear layer to obtain the predicted blood pressure value. The model is trained with the Adam optimizer (learning rate 0.01) for 1000 epochs using MSE loss.

## C. Detailed blood pressure estimation performance for all models across subjects and activities

Table C.1 provided the detailed metrics value for systolic and diastolic pressure estimation from each model across all subject for each activity. The percentages of samples whose error falls within the AAMI standard (± 5 mmHg) are provided in the second row, while the median, minimum and maximum value of 30 values in each column are listed at the bottom. Statistical comparison results between PMB-NN

and the benchmark models for each activity are provided as p and d values. Significant p and d values are bolded when PMB-NN outperforms the reference model, and bolded with an underline when PMB-NN underperforms.

## D. Correlation analysis results for estimated $P_s$-$T_s$ and estimated $P_d$-$T_d$

Table D.1 provided the Spearman's correlation coefficient ($\rho$) between the estimated $P_s$ and input $T_s$, between the estimated $P_d$ and input $T_d$ from each model for each data segment. Median, min and max values are summarized, while statistical analysis was performed to compare PMB-NN and other four models.

## CRediT authorship contribution statement

**Yaowen Zhang:** Conceptualization, Data curation, Formal Analysis, Methodology, Software, Validation, Visualization, Writing - Original Draft, Writing – review and editing. **Libera Fresiello:** Methodology, Supervision, Writing – review and editing. **Peter H. Veltink:** Supervision, Writing - Review and Editing. **Dirk W. Donker:** Supervision, Writing - Review and Editing. **Ying Wang:** Conceptualization, Methodology, Resources, Supervision, Writing - Review and Editing.

## Statements of ethical approval

The authors declare that the data used in this work was collected according to the guidelines of the Declaration of Helsinki, approved by the Ethics Committee for Computer and Information Science (EC-CIS) of the University of Twente (Approval No. 240831) and conducted at Roessingh Research and Development (RRD).

## Funding

The current research was partly funded by the China Scholarship Council (CSC).

## Declaration of competing interest

The authors have no competing interests to declare.

| | PMB-NN | | | | | | FCNN | | | | | | CNN-LSTM | | | | | | Transformer | | | | | | PM | | | | | |
|---|---|---|---|---|---|---|---|---|---|---|---|---|---|---|---|---|---|---|---|---|---|---|---|---|---|---|---|---|---|---|---|
| ±5 mmHg | $P_s$: 39.32% | | | $P_d$: 69.09% | | | $P_s$: 39.87% | | | $P_d$: 74.93% | | | $P_s$: 40.32% | | | $P_d$: 75.80% | | | $P_s$: 32.43% | | | $P_d$: 65.17% | | | $P_s$: 32.28% | | | $P_d$: 58.11% | | |
| | ME | SD | MAE | ME | SD | MAE | ME | SD | MAE | ME | SD | MAE | ME | SD | MAE | ME | SD | MAE | ME | SD | MAE | ME | SD | MAE | ME | SD | MAE | ME | SD | MAE |
| *Static Postures (SP)* | | | | | | | | | | | | | | | | | | | | | | | | | | | | | | |
| 01-SP | -0.39 | 6.62 | 5.35 | -4.4 | 4.28 | 5.63 | -0.61 | 6.61 | 5.35 | -1.85 | 3.3 | 3.17 | -0.6 | 6.64 | 5.35 | -1.57 | 3.23 | 2.92 | 0.78 | 9.84 | 7.71 | -0.79 | 6.25 | 4.87 | 21.92 | 11.08 | 22.32 | 7.66 | 6.84 | 8.54 |
| 02-SP | 4.06 | 7.36 | 7.02 | -4.39 | 6.22 | 6.45 | 4.19 | 7.42 | 7.11 | -1.92 | 5.33 | 4.22 | 4.59 | 7.34 | 7.27 | -1.56 | 5.22 | 4.04 | 6.7 | 14.42 | 12.88 | -1.37 | 9.32 | 7.06 | 4.06 | 7.08 | 6.69 | -2.28 | 7.05 | 5.40 |
| 03-SP | 4.55 | 7.09 | 6.81 | -0.51 | 7.03 | 3.87 | 4.37 | 7.09 | 6.68 | 0.25 | 6.79 | 3.28 | 4.22 | 6.92 | 6.58 | -0.66 | 6.96 | 3.94 | 3.64 | 11.9 | 9.08 | -0.78 | 6.22 | 4.59 | -4.52 | 7.15 | 6.25 | -3.82 | 6.56 | 6.17 |
| 04-SP | 1.86 | 8.3 | 6.95 | 2.77 | 3.18 | 3.38 | 2.05 | 8.32 | 7.03 | 3.47 | 3.38 | 4.15 | 2.49 | 8.26 | 7.19 | 3.14 | 4.32 | 4.57 | 1.96 | 11.4 | 9.24 | 2.55 | 4.76 | 4.27 | 0.64 | 10.13 | 7.76 | 2.09 | 3.66 | 3.35 |
| 05-SP | 4.43 | 6.85 | 6.94 | -1.52 | 3.67 | 2.83 | 4.89 | 6.83 | 7.22 | -0.83 | 3.32 | 2.28 | 4.73 | 6.84 | 7.12 | -0.75 | 3.23 | 2.12 | 3.07 | 10.01 | 7.94 | -1.06 | 5.88 | 4.67 | -1.10 | 6.09 | 4.64 | -2.41 | 4.48 | 4.06 |
| 06-SP | 1.06 | 8.82 | 6.27 | 2.13 | 4.07 | 3.29 | 1.3 | 8.9 | 6.41 | 3.89 | 4.01 | 4.44 | -1.97 | 8.64 | 5.79 | 2.37 | 4.94 | 4.26 | 2.58 | 14.85 | 9.79 | 4.11 | 5.76 | 5.16 | -0.98 | 10.94 | 7.56 | 2.81 | 6.59 | 4.82 |
| 07-SP | -4.28 | 9.02 | 7.43 | -1.1 | 5.52 | 4.05 | -4.09 | 9.06 | 7.4 | -0.25 | 5.04 | 3.49 | -3.82 | 8.81 | 7.15 | -0.09 | 5.21 | 3.39 | -4.95 | 11.73 | 10.06 | -0.39 | 5.22 | 4.00 | -5.49 | 9.32 | 8.09 | 0.69 | 5.24 | 3.75 |
| 08-SP | 1.09 | 7.14 | 5.84 | -1.13 | 7.76 | 5.55 | 1.9 | 7.3 | 6.12 | 0.99 | 5.04 | 3.41 | 1.57 | 7.06 | 5.83 | 1.08 | 5.55 | 3.55 | 0.22 | 10.68 | 8.59 | 0.72 | 6.3 | 3.65 | -11.37 | 7.42 | 11.56 | -8.18 | 5.06 | 8.92 |
| 09-SP | -1.86 | 8.66 | 5.87 | 2.15 | 5.36 | 5.02 | -1.62 | 8.69 | 5.95 | 2.07 | 4.26 | 3.21 | -1.43 | 8.65 | 5.96 | 1.78 | 3.95 | 2.98 | -1.01 | 6.79 | 5.19 | 1.39 | 4.99 | 3.64 | -2.53 | 10.03 | 6.34 | 1.41 | 10.10 | 4.85 |
| 10-SP | 5.42 | 9.12 | 9.09 | -3.61 | 5.08 | 5 | 5.38 | 9.06 | 9.01 | -0.99 | 4.47 | 3.33 | 5.15 | 9.77 | 9.48 | -1.81 | 3.77 | 3.13 | 2.39 | 9.62 | 8.2 | -2.72 | 4.19 | 3.91 | 6.10 | 9.57 | 9.63 | -0.58 | 5.73 | 4.11 |
| p-value | – | – | – | – | – | – | 0.275 | 0.123 | 0.137 | 0.131 | 0.006 | 0.105 | 0.322 | 0.160 | 0.859 | 0.084 | 0.105 | 0.131 | 0.695 | 0.006 | 0.010 | 0.084 | 0.232 | 0.695 | 0.557 | 0.049 | 0.160 | 0.557 | 0.131 | 0.084 |
| d-value | – | – | – | – | – | – | -0.079 | -0.031 | -0.069 | 0.545 | **0.555** | 0.897 | -0.091 | 0.005 | -0.013 | 0.764 | 0.435 | 0.873 | 0.089 | **-1.789** | **-1.340** | 0.609 | -0.467 | -0.195 | -0.624 | **-0.682** | -0.641 | -0.385 | -0.560 | -0.639 |
| *Low Intensity Cycling (LIC)* | | | | | | | | | | | | | | | | | | | | | | | | | | | | | | |
| 01-LIC | -1.94 | 7.38 | 5.89 | 2.58 | 3.79 | 3.87 | -1.83 | 7.37 | 5.87 | 2.97 | 2.95 | 3.41 | -1.66 | 7.24 | 5.76 | 2.83 | 2.88 | 3.31 | -1.13 | 11.85 | 9.13 | 1.77 | 3.25 | 3.06 | 4.71 | 7.55 | 7.42 | 7.75 | 3.26 | 7.77 |
| 02-LIC | -4.67 | 12.19 | 10.35 | 3.02 | 5.25 | 5.16 | -5.31 | 12.14 | 10.41 | 3.68 | 5.37 | 5.39 | -4.18 | 12.23 | 10.35 | 3.2 | 5.62 | 5.36 | -3.87 | 18.39 | 15.25 | 2.89 | 8.3 | 7.08 | -8.25 | 12.98 | 12.10 | 2.23 | 6.69 | 5.90 |
| 03-LIC | 0.79 | 5.15 | 4.09 | 0.39 | 6.17 | 3.71 | 1.34 | 5.15 | 4.19 | 2.63 | 2.51 | 3.14 | 1.44 | 5.15 | 4.2 | 2.37 | 2.81 | 3.12 | 2.04 | 9.61 | 7.84 | 2.16 | 4.07 | 3.80 | -2.10 | 4.69 | 4.05 | 2.33 | 5.59 | 4.56 |
| 04-LIC | -4.57 | 7.95 | 6.4 | 0.5 | 3.03 | 2.45 | -3.9 | 7.95 | 6.18 | -0.65 | 3.14 | 2.45 | -3.35 | 7.73 | 5.77 | -0.38 | 3.02 | 2.32 | -2.66 | 9.76 | 8.19 | -1.44 | 3.67 | 3.32 | -6.02 | 7.81 | 6.94 | 0.07 | 3.36 | 2.40 |
| 05-LIC | -0.91 | 7.52 | 5.92 | -0.04 | 3.83 | 2.71 | -0.62 | 7.48 | 5.89 | 0.89 | 4.73 | 3.41 | 0.43 | 6.83 | 5.33 | 0.02 | 3.62 | 2.6 | 1.8 | 10.27 | 8.36 | -0.6 | 3.56 | 2.94 | -2.88 | 6.92 | 5.50 | -0.75 | 3.36 | 2.40 |
| 06-LIC | 8.37 | 10.64 | 11.35 | -2.54 | 3.38 | 3.27 | 8.38 | 10.64 | 11.35 | -2.12 | 3.44 | 3.07 | 8.01 | 10.6 | 11.13 | -0.94 | 3.3 | 3.44 | 7.18 | 14.02 | 12.52 | -1.82 | 3.68 | 3.23 | 5.51 | 10.71 | 10.05 | -2.23 | 4.25 | 3.34 |
| 07-LIC | -1.9 | 7.34 | 6.06 | -1.44 | 5.29 | 3.9 | -1.57 | 7.51 | 6.01 | 0.66 | 5.31 | 3.62 | -1.58 | 7.31 | 6.06 | 0.42 | 5.15 | 3.90 | -0.33 | 18.11 | 15.01 | -0.11 | 7.04 | 4.62 | -0.31 | 9.94 | 7.70 | -0.51 | 6.44 | 4.55 |
| 08-LIC | -7.85 | 12.65 | 12.49 | 2.39 | 3.93 | 3.78 | -6.76 | 14.21 | 13.44 | 3.11 | 3.65 | 3.99 | -4.83 | 9.46 | 8.49 | 3.05 | 3.66 | 4.18 | -5.52 | 10.96 | 10.09 | 2.58 | 3.75 | 3.68 | -16.17 | 12.75 | 17.41 | 2.05 | 4.72 | 4.02 |
| 09-LIC | -4.25 | 6.36 | 6.4 | 1.35 | 6.02 | 5.02 | -4.64 | 6.85 | 6.99 | 2.13 | 4.87 | 4.34 | -4.3 | 8.65 | 6.65 | 1.9 | 4.79 | 4.18 | -2.46 | 9.59 | 8.35 | 0.65 | 5.08 | 3.83 | -11.53 | 7.80 | 11.78 | -2.21 | 6.39 | 4.62 |
| 10-LIC | 7.22 | 7.86 | 8.27 | -0.18 | 6.29 | 4.22 | 7.81 | 7.86 | 8.65 | 2.17 | 4.19 | 3.69 | 7.43 | 7.77 | 8.45 | 2.12 | 3.45 | 3.09 | 17.38 | 9.87 | 17.58 | 3.71 | 4.35 | 4.65 | 0.20 | 15.31 | 11.08 | -2.54 | 8.86 | 5.68 |
| p-value | – | – | – | – | – | – | 1.000 | 0.345 | 0.214 | 0.084 | 0.322 | 0.314 | 0.105 | 0.123 | 0.208 | 0.375 | **0.027** | **0.020** | 0.375 | 0.004 | 0.014 | 0.770 | 0.846 | 0.492 | 0.322 | 0.193 | 0.037 | 0.375 | 0.131 | 0.084 |
| d-value | – | – | – | – | – | – | 0.011 | -0.081 | -0.061 | -0.6 | 0.601 | 0.189 | 0.194 | 0.177 | 0.198 | -0.241 | **0.772** | **0.475** | -0.047 | **-1.245** | **-1.096** | -0.293 | 0.016 | -0.199 | -0.37 | -0.392 | **-0.497** | -0.485 | -0.357 | -0.542 |
| *Moderate Intensity Cycling (MIC)* | | | | | | | | | | | | | | | | | | | | | | | | | | | | | | |
| 01-MIC | 8.34 | 6.82 | 9.49 | -0.72 | 2.62 | 1.87 | 7.84 | 6.89 | 9.14 | -0.5 | 2.16 | 1.65 | 7.69 | 6.89 | 9.03 | -1.43 | 2.04 | 2.11 | 7.38 | 9.84 | 9.93 | -1.73 | 4.73 | 4.04 | 8.59 | 6.99 | 9.78 | 0.93 | 4.11 | 2.81 |
| 02-MIC | -1.84 | 11.29 | 8.66 | 3.97 | 5.64 | 5.15 | -1.65 | 11.28 | 8.66 | 4.62 | 5.59 | 5.5 | -1.69 | 11.38 | 8.67 | 3.19 | 5.57 | 4.67 | -0.58 | 17.3 | 12.73 | 2.94 | 10.3 | 7.94 | -12.51 | 10.73 | 13.23 | -1.32 | 5.80 | 4.29 |
| 03-MIC | -8.86 | 6.54 | 9.46 | -3.7 | 2.9 | 4.03 | -4.6 | 6.44 | 6.39 | 0.26 | 3.48 | 2.54 | -7.06 | 6.57 | 8.2 | 0.41 | 3.53 | 2.56 | -8.59 | 12.2 | 11.36 | -1.99 | 6.37 | 5.39 | -9.81 | 7.27 | 10.88 | -6.30 | 4.49 | 6.53 |
| 04-MIC | -1.89 | 5.84 | 4.34 | -4.75 | 3.12 | 5.06 | -1.92 | 5.84 | 4.35 | -3.47 | 2.99 | 3.91 | -2.15 | 5.87 | 4.49 | -3.17 | 2.8 | 3.59 | -2.45 | 8.28 | 6.76 | -4.03 | 4.09 | 4.84 | -9.86 | 6.67 | 10.06 | -5.41 | 4.61 | 5.85 |
| 05-MIC | 4.18 | 10.6 | 9.12 | -2.21 | 4.08 | 3.64 | 3.38 | 10.5 | 8.94 | -1.82 | 3.95 | 3.43 | 4.95 | 10.15 | 9.2 | -2.15 | 4.21 | 3.77 | 5.45 | 14.14 | 12.64 | -2.95 | 5.19 | 4.74 | -8.46 | 11.10 | 11.18 | -8.33 | 6.15 | 8.71 |
| 06-MIC | 2.81 | 9.75 | 11.42 | 2.45 | 3.84 | 3.36 | 2.86 | 11.56 | 9.94 | 2.88 | 3.68 | 3.58 | 2.89 | 11.47 | 9.84 | 2.88 | 3.67 | 3.57 | 3.37 | 15.18 | 12.3 | 0.53 | 4.8 | 3.82 | -0.67 | 12.27 | 9.76 | 0.43 | 4.46 | 3.34 |
| 07-MIC | -6.62 | 11.54 | 10.95 | 0.62 | 7.54 | 6.06 | -5.85 | 11.9 | 10.84 | 1.78 | 5.65 | 4.53 | -5.96 | 11.61 | 10.7 | 1.9 | 5.66 | 4.57 | -5.07 | 15.01 | 13.17 | 1.11 | 6.36 | 4.90 | 14.82 | 11.80 | 15.97 | 11.68 | 8.48 | 12.92 |
| 08-MIC | 2.95 | 13.17 | 10.03 | -0.74 | 7.07 | 5.19 | 3.75 | 13.93 | 10.82 | -0.29 | 6.21 | 4.53 | 3.76 | 15.01 | 11.67 | -1.22 | 5.82 | 4.50 | 4.5 | 16.51 | 13.81 | -2.38 | 5.49 | 4.82 | -16.04 | 12.29 | 17.73 | -8.53 | 7.58 | 9.67 |
| 09-MIC | -1.25 | 14.62 | 12.28 | -2.73 | 4 | 3.88 | -1.27 | 14.85 | 12.47 | -1.6 | 4.3 | 3.53 | -1.65 | 13.57 | 11.2 | -1.94 | 3.71 | 3.30 | -1.1 | 10.89 | 8.64 | -2.71 | 3.75 | 3.88 | -5.32 | 13.93 | 11.89 | -2.94 | 5.05 | 4.16 |
| 10-MIC | 3.43 | 16.98 | 15.01 | -1.3 | 6.63 | 5.08 | 4.58 | 17.28 | 15.31 | 0.66 | 4.61 | 3.67 | 5.24 | 16.24 | 14.91 | 0.19 | 4.33 | 3.46 | 12.08 | 11.33 | 12.88 | 1.98 | 4.41 | 3.88 | -11.39 | 32.35 | 29.04 | -7.00 | 14.27 | 12.79 |
| p-value | – | – | – | – | – | – | 0.695 | 0.086 | 0.859 | 0.322 | 0.131 | 0.037 | 0.557 | 0.770 | 0.557 | 0.322 | 0.064 | **0.027** | 0.625 | 0.193 | 0.105 | 0.770 | 0.492 | 0.314 | **0.010** | 0.375 | 0.010 | 0.084 | 0.002 | **0.020** |
| d-value | – | – | – | – | – | – | 0.184 | -0.045 | 0.077 | 0.359 | 0.306 | **0.563** | -0.035 | 0.002 | 0.044 | 0.366 | 0.389 | **0.692** | -0.265 | -0.657 | -0.604 | 0.066 | -0.434 | -0.405 | **-1.487** | -0.284 | **-0.875** | -1.034 | -0.693 | **-0.996** |
| *Overall* | | | | | | | | | | | | | | | | | | | | | | | | | | | | | | |
| Median | 0.2 | 8.13 | 7.23 | -0.62 | 4.68 | 3.89 | 0.35 | 8.14 | 7.17 | 0.66 | 4.23 | 3.51 | -0.08 | 8.02 | 7.23 | 0.30 | 3.86 | 3.45 | 1.29 | 11.37 | 9.86 | -0.25 | 5.04 | 4.15 | -2.70 | 9.98 | 9.92 | -0.66 | 5.66 | 4.72 |
| Max. | 8.37 | 16.98 | 15.01 | 3.97 | 7.76 | 6.45 | 8.38 | 17.28 | 15.31 | 4.62 | 6.79 | 5.50 | 8.01 | 16.24 | 14.91 | 3.20 | 6.96 | 5.36 | 17.38 | 18.39 | 17.58 | 4.11 | 10.30 | 7.94 | 21.92 | 32.35 | 29.04 | 11.68 | 14.27 | 12.92 |
| Min. | -8.86 | 5.15 | 4.09 | -4.75 | 2.62 | 1.87 | -6.76 | 5.15 | 4.19 | -3.47 | 2.16 | 1.65 | -7.06 | 5.15 | 4.20 | -3.17 | 2.04 | 2.11 | -8.59 | 6.79 | 5.19 | -4.03 | 3.25 | 2.94 | -16.17 | 4.69 | 4.05 | -8.53 | 3.07 | 2.40 |
| p-value | – | – | – | – | – | – | **1.000** | 0.016 | 0.149 | 0.626 | **0.001** | **0.004** | 0.824 | 0.144 | 0.254 | 0.253 | **0.001** | **0.000** | 0.477 | **0.000** | **0.000** | 0.428 | 0.221 | 0.198 | **0.004** | 0.010 | **0.000** | 0.067 | **0.000** | **0.001** |
| d-value | – | – | – | – | – | – | 0.048 | **-0.047** | -0.003 | 0.151 | **0.476** | **0.571** | 0.040 | 0.050 | 0.079 | 0.298 | **0.513** | **0.697** | -0.090 | **-1.051** | **-0.869** | 0.145 | -0.304 | -0.271 | **-0.776** | **-0.316** | **-0.638** | -0.639 | **-0.556** | **-0.720** |

Table C.1: Detailed blood pressure estimation performance (systolic ($P_s$) and diastolic ($P_d$)) for all models across subjects and activities.

| Subject | Estimated $P_s$ vs $T_s$ | | | | | Estimated $P_d$ vs $T_d$ | | | | |
|---|---|---|---|---|---|---|---|---|---|---|
| | PMB-NN | FCNN | CNN-LSTM | Transformer | PM | PMB-NN | FCNN | CNN-LSTM | Transformer | PM |
| *Static Postures (SP)* | | | | | | | | | | |
| 01-SP | -1.000 | -1.000 | -1.000 | 0.099 | 0.191 | -1.000 | 0.914 | 0.959 | -0.352 | -0.856 |
| 02-SP | -1.000 | -1.000 | -1.000 | -0.140 | -0.104 | -1.000 | 1.000 | 1.000 | -0.079 | -0.917 |
| 03-SP | -1.000 | -1.000 | -1.000 | 0.217 | 0.518 | -1.000 | -1.000 | 0.640 | -0.027 | -0.639 |
| 04-SP | -1.000 | -1.000 | -1.000 | -0.005 | -0.241 | -1.000 | -1.000 | -0.240 | -0.281 | -0.947 |
| 05-SP | -1.000 | -1.000 | -1.000 | -0.084 | -0.247 | -1.000 | 0.626 | -1.000 | 0.021 | -0.925 |
| 06-SP | -1.000 | -1.000 | -0.810 | 0.148 | -0.452 | -0.119 | -1.000 | 0.807 | -0.400 | 0.001 |
| 07-SP | -1.000 | -1.000 | -1.000 | -0.133 | -0.141 | -1.000 | -1.000 | -0.787 | -0.433 | -0.922 |
| 08-SP | -1.000 | -1.000 | -1.000 | -0.123 | 0.140 | -1.000 | 0.702 | 0.710 | -0.272 | -0.915 |
| 09-SP | -1.000 | -1.000 | -1.000 | 0.020 | 0.275 | -1.000 | -1.000 | -0.917 | 0.050 | -0.804 |
| 10-SP | -1.000 | -1.000 | 1.000 | -0.102 | -0.134 | -1.000 | -1.000 | -1.000 | -0.169 | -0.753 |
| p-value | – | 1.000 | 0.180 | **0.002** | **0.002** | – | 0.080 | **0.012** | **0.004** | **0.002** |
| d-value | – | 0.000 | -0.438 | **-109.354** | **-45.403** | – | -0.492 | **-1.299** | **-2.777** | **-2.866** |
| *Low Intensity Cycling (LIC)* | | | | | | | | | | |
| 01-LIC | -1.000 | -1.000 | -1.000 | -0.236 | 0.388 | -1.000 | -1.000 | -1.000 | -0.231 | -0.570 |
| 02-LIC | -1.000 | -1.000 | -1.000 | -0.190 | -0.721 | -1.000 | -1.000 | -0.652 | 0.008 | -0.890 |
| 03-LIC | -1.000 | -1.000 | -1.000 | 0.216 | -0.398 | -1.000 | -1.000 | -1.000 | 0.053 | -0.915 |
| 04-LIC | -1.000 | -1.000 | -1.000 | 0.303 | 0.137 | -1.000 | -1.000 | -0.997 | -0.239 | -0.935 |
| 05-LIC | -1.000 | -1.000 | -1.000 | 0.049 | -0.133 | -1.000 | -1.000 | -1.000 | -0.291 | -0.953 |
| 06-LIC | -1.000 | -1.000 | -1.000 | 0.097 | -0.139 | -1.000 | -1.000 | -1.000 | -0.051 | -0.984 |
| 07-LIC | -1.000 | -1.000 | -1.000 | -0.123 | -0.279 | -1.000 | 1.000 | -1.000 | -0.025 | -0.664 |
| 08-LIC | -1.000 | -1.000 | 0.999 | 0.200 | -0.654 | -1.000 | -1.000 | -1.000 | -0.009 | -0.729 |
| 09-LIC | -1.000 | -1.000 | -1.000 | -0.034 | 0.164 | -1.000 | -1.000 | -1.000 | -0.236 | -0.946 |
| 10-LIC | -1.000 | -1.000 | -1.000 | -0.031 | -0.527 | -1.000 | -1.000 | -1.000 | 0.030 | -0.800 |
| p-value | – | 1.000 | 0.317 | **0.002** | **0.002** | – | 0.317 | 0.180 | **0.002** | **0.002** |
| d-value | – | 0.000 | -0.316 | **-77.136** | **-32.526** | – | -0.316 | -0.471 | **-103.274** | **-23.095** |
| *Moderate Intensity Cycling (MIC)* | | | | | | | | | | |
| 01-MIC | -1.000 | -1.000 | -1.000 | -0.301 | -0.418 | -1.000 | -1.000 | 0.950 | -0.170 | -0.899 |
| 02-MIC | -1.000 | -1.000 | -1.000 | -0.031 | -0.767 | -1.000 | -1.000 | -1.000 | -0.242 | -0.818 |
| 03-MIC | -1.000 | -0.392 | -1.000 | 0.249 | 0.762 | -1.000 | 1.000 | 1.000 | -0.230 | 0.381 |
| 04-MIC | -1.000 | -1.000 | -1.000 | -0.235 | -0.825 | -1.000 | -1.000 | -1.000 | -0.205 | -0.918 |
| 05-MIC | -1.000 | -1.000 | -1.000 | -0.010 | -0.785 | -1.000 | -1.000 | -1.000 | 0.049 | -0.968 |
| 06-MIC | -1.000 | -1.000 | -1.000 | 0.135 | -0.409 | -1.000 | -1.000 | -1.000 | -0.061 | -0.941 |
| 07-MIC | -1.000 | -1.000 | -1.000 | -0.032 | 0.243 | -1.000 | 0.999 | 1.000 | -0.002 | -0.890 |
| 08-MIC | -1.000 | -1.000 | -1.000 | -0.546 | -0.571 | -1.000 | -1.000 | -1.000 | -0.441 | -0.935 |
| 09-MIC | -1.000 | -1.000 | -1.000 | 0.002 | -0.425 | -1.000 | -1.000 | -1.000 | -0.589 | -0.939 |
| 10-MIC | -1.000 | -1.000 | -1.000 | 0.349 | -0.682 | -1.000 | -0.691 | -1.000 | 0.017 | -0.354 |
| p-value | – | 0.317 | 1.000 | **0.002** | **0.002** | – | 0.109 | 0.103 | **0.002** | **0.002** |
| d-value | – | -0.316 | 0.000 | **-49.757** | **-20.579** | – | -0.581 | -0.597 | **-60.879** | **-17.254** |
| *Overall* | | | | | | | | | | |
| Median | -1.000 | -1.000 | -1.000 | -0.021 | -0.244 | -1.000 | -1.000 | -1.000 | -0.170 | -0.907 |
| Max | -1.000 | -1.000 | 1.000 | 0.349 | 0.762 | -0.119 | 1.000 | 1.000 | 0.053 | 0.381 |
| Min | -1.000 | -1.000 | -1.000 | -0.456 | -0.825 | -1.000 | -1.000 | -1.000 | -0.589 | -0.984 |
| p-value | – | 0.317 | **0.018** | **0.000** | **0.000** | – | **0.015** | **0.000** | **0.000** | **0.000** |
| d-value | – | -0.183 | **-0.463** | **-69.335** | **-27.230** | – | **-0.477** | **-0.890** | **-5.166** | **-5.201** |

Table D.1: Spearman correlation coefficients (sorted by activity)